\documentclass[journal]{IEEEtran}
\newcommand{\RNum}[1]{\uppercase\expandafter{\romannumeral #1\relax}}
\usepackage{mathrsfs}
\usepackage{amssymb}
\usepackage[mathcal]{euscript}
\usepackage{diagbox}
\usepackage{cite}
\usepackage[T1]{fontenc}
\usepackage{graphicx}
\usepackage{CJKutf8}
\usepackage{makecell}
\usepackage{psfrag}
\usepackage{url}
\usepackage{stfloats}
\usepackage{amsmath}
\usepackage{array}
\usepackage{float}
\usepackage{multirow} 
\usepackage{hyperref}
\usepackage{amsthm}
\usepackage{color}
\usepackage{multicol}
\usepackage{stfloats}
\usepackage{enumerate}
\usepackage{subfigure}
\usepackage{booktabs}  
\newtheorem{remark}{Remark}

\newtheorem{example}{Example}

\allowdisplaybreaks[4]
\usepackage[ruled]{algorithm2e}
\usepackage{algorithmic} 
\usepackage{xurl}

\ifCLASSINFOpdf
\else
\fi
\begin{document}
\begin{CJK}{UTF8}{gbsn}
%
\title{Compact LLM Deployment and World Model Assisted Offloading in Mobile Edge Computing}



%
	\author{Ruichen Zhang, Xiaofeng Luo, Jiayi He, Jiawen Kang,  Zehui Xiong, Shiwen Mao,~\IEEEmembership{Fellow,~IEEE}

\thanks{R. Zhang is with the College of Computing and Data Science, Nanyang Technological University, Singapore (e-mail: ruichen.zhang@ntu.edu.sg).}

\thanks{X. Luo, J. He, and J. Kang are with the School of Automation,
Guangdong University of Technology, Guangzhou 510006, China (e-mail: gdutxiaofengluo@163.com, jiayihe@ieee.org, kavinkang@gdut.edu.cn).}

\thanks{Z. Xiong is with the School of Electronics, Electrical Engineering and
Computer Science, Queen’s University Belfast, BT9 5BN Belfast, U.K. (email: z.xiong@qub.ac.uk). }

\thanks{S. Mao is with the Department of Electrical and Computer Engineering, Auburn University, USA (e-mail: smao@ieee.org).}

}
\maketitle

\begin{abstract}
This paper investigates compact large language model (LLM) deployment and world-model-assisted inference offloading in mobile edge computing (MEC) networks. We first propose an edge compact LLM deployment (ECLD) framework that jointly applies structured pruning, low-bit quantization, and knowledge distillation to construct edge-deployable LLM variants, and we evaluate these models using four complementary metrics: accessibility, energy consumption, hallucination rate, and generalization accuracy. Building on the resulting compact models, we formulate an MEC offloading optimization problem that minimizes the long-term average inference latency subject to per-device energy budgets and LLM-specific quality-of-service constraints on effective accuracy and hallucination. To solve this problem under unknown and time-varying network dynamics, we develop a world model-proximal policy optimization (PPO) algorithm, which augments an on-policy PPO algorithm with a learned recurrent world model that provides improved value targets and short imagination rollouts. Extensive experiments on Llama-3.1-8B, Qwen3-8B, and Mistral-12B show that ECLD compresses base models by about 70–80\% in storage (i.e., from 15.3 GB to 3.3 GB for Llama-3.1-8B) and reduces per-query energy consumption by up to 50\%, while largely preserving accuracy and often lowering hallucination compared with quantization-only or pruning-only baselines. Moreover, they also show that world model-PPO speeds up convergence by about 50\%, improves the final reward by 15.8\% over vanilla PPO, and reduces average inference latency by 12–30\% across different user populations, while satisfying the accuracy and hallucination constraints and approaching the generation quality of always-offloading with much of the efficiency of local execution.
\end{abstract}

\begin{IEEEkeywords}
Edge LLM, MEC, compact model, world model,  Offloading, PPO.
\end{IEEEkeywords}

\section{Introduction}

The recent rise of generative artificial intelligence (GAI) has led to the rapid development of large language models (LLMs), such as OpenAI's ChatGPT and DeepSeek's R1, which demonstrate remarkable capabilities in text generation, reasoning, and multimodal understanding~\cite{zhang2024generative,liu2024deep}. These frontier models are typically of GPT-3 scale or larger, with on the order of $10^{11}$–$10^{12}$ parameters~\cite{brown2020language}. For instance, a 175B-parameter model requires hundreds of gigabytes of memory in FP16 merely to store the weights, while smaller open-source models with $7$–$13$\,B parameters still demand tens of gigabytes in standard precision. In contrast, mainstream smartphones and edge devices usually provide only $4$–$12$\,GB of RAM and operate under strict power and thermal limitations, making direct deployment of such LLMs on end devices infeasible~\cite{zhang2025toward}. As a result, most commercial services still rely on cloud-hosted LLMs, with user prompts being transmitted from edge devices to remote data centers~\cite{qu2025mobile}. Although this cloud-centric architecture can deliver strong performance, it exhibits two inherent drawbacks: (\emph{i}) The end-to-end quality of experience is dominated by wide-area network conditions, where round-trip times of hundreds of milliseconds are common even before model computation~\cite{qu2025mobile}. (\emph{ii}) The service remains tightly coupled to backhaul connectivity, which is undesirable for latency-sensitive or intermittently connected edge applications~\cite{zhang2024beyond}.

To alleviate these limitations, there is a growing interest in {compact LLMs} that can be deployed locally on mobile phones, tablets, and edge nodes~\cite{qu2025mobile}. By pruning redundant parameters~\cite{ma2023llm}, distilling knowledge from larger teacher models~\cite{yang2025survey}, and quantizing weights to low-bit precision~\cite{xiao2023smoothquant}, compact LLMs can reduce the memory footprint from tens of gigabytes to a few gigabytes and substantially lower the per-token computation cost, while still preserving acceptable task performance on many everyday workloads. Nevertheless, purely on-device inference is often insufficient in practice: local compact LLMs may still struggle with complex instructions, domain-specific queries, or long-context reasoning, and their behavior (e.g., accuracy and hallucination rate) is highly sensitive to resource constraints and quantization levels. This motivates a collaborative {mobile edge computing} (MEC) paradigm, in which each user device runs a lightweight LLM locally and selectively offloads part of the inference workload to an edge server hosting a full-scale LLM~\cite{10591707, zhang2025toward}. In this setting, the key question becomes how to jointly design compact LLM deployment and edge-assisted offloading so that users can enjoy low-latency, energy-efficient, and trustworthy LLM services. However, realizing such an edge-assisted compact LLM system faces several fundamental challenges, as elaborated below.

\textbf{\uppercase\expandafter{\romannumeral1}). Compact LLM Deployment Under Diverse Device Capabilities:}
Mobile and edge devices span a wide range of CPU configurations, memory capacities, and energy budgets: for example, high-end smartphones may provide several performance cores and $8$--$12$\,GB of RAM, whereas low-end handsets or IoT nodes often operate with only a few gigabytes of memory and tight battery constraints~\cite{shuvo2022efficient, 10596048}. Designing a single compression configuration for all devices is therefore suboptimal: aggressive pruning and quantization can severely degrade accuracy or exacerbate hallucinations on resource-limited platforms, while conservative compression wastes memory and compute on more capable devices and limits the achievable latency reduction~\cite{zhang2024edgeshard, zheng2025review}. This calls for a {systematic} deployment framework that jointly coordinates pruning, knowledge distillation, and quantization in a hardware-aware manner. The key challenge is to co-design these steps in a unified pipeline so that the resulting compact LLMs remain lightweight enough for on-device deployment yet accurate enough to support end-to-end LLM services.

\textbf{\uppercase\expandafter{\romannumeral2}). MEC Offloading and QoS-Aware Adaptive Resource Control:}
Compact-LLM–enabled MEC systems also operate under highly dynamic and diverse conditions in terms of traffic arrivals, wireless channels, and user-level quality-of-service (QoS) requirements. For instance, bursty inference requests from multiple users must be served over interference-limited uplinks, while different applications may impose distinct latency, energy, or content-fidelity targets~\cite{8115180}. Designing a fixed offloading or power-control rule for all scenarios is therefore suboptimal: aggressive local execution on compact models can violate accuracy or hallucination constraints when tasks are complex or device resources are scarce~\cite{lee2024exploring}, whereas always offloading to the edge server improves QoS but incurs excessive transmission delay and uplink power consumption~\cite{zhang2024beyond}. This motivates a {QoS-aware} adaptive control framework that jointly tunes offloading ratios and transmit powers based on real-time network states while explicitly accounting for application-level QoS and environment uncertainty. The key challenge is to learn such a control policy directly from interaction with the MEC environment in a sample-efficient and stable manner, motivating a world-model–enhanced design that plans over imagined trajectories rather than reacting myopically to instantaneous feedback.

To tackle these challenges, two important families of techniques have emerged, i.e., LLM compression and edge deployment and world model-reinforcement learning for adaptive MEC control.

\textbf{\uppercase\expandafter{\romannumeral1}). LLM Compression and Edge Deployment:}
Recent studies on LLM compression have explored several model–shrinking techniques, including structured pruning, low-rank decomposition, knowledge distillation, and low-bit quantization~\cite{frantar2023sparsegpt,dettmers2023qlora}. Structured pruning removes redundant channels, attention heads, and feed-forward blocks according to saliency scores~\cite{ma2023llm}. In this way, the transformer architecture is preserved and can still be implemented efficiently on commodity hardware. Low-rank and adapter-based methods factorize weight matrices or insert trainable low-rank updates, which reduces the effective parameter count without fully retraining the backbone~\cite{zanella2024low}. Knowledge distillation transfers soft logits or hidden representations from a large teacher to a smaller student model and compensates for the capacity loss induced by pruning~\cite{yang2025survey}. Quantization further compresses the model by representing weights and activations with mixed-precision formats (e.g., 8-bit or 4-bit), often combined with outlier-aware schemes and KV-cache quantization to accelerate generation~\cite{xiao2023smoothquant}. These techniques can greatly reduce model size and computation, enabling deployment on consumer-grade GPUs or high-end smartphones. However, most existing schemes are designed in isolation for a specific hardware platform or benchmark. They do not provide a unified framework that couples pruning, distillation, quantization, and hardware-aware deployment for diverse MEC environments. In addition, they usually optimize standard NLP metrics on static datasets, without explicitly considering how the compressed LLM will interact with an edge-assisted offloading system and its latency–energy–QoS trade-offs.

\textbf{\uppercase\expandafter{\romannumeral2}). World–Model–Enhanced Reinforcement Learning:}
Deep reinforcement learning (DRL) has been widely used in MEC to optimize task offloading, resource allocation, and energy management~\cite{mao2018deep,chen2021drl}. These methods can adapt to time-varying wireless environments and user demands. However, classical DRL algorithms are mostly {reactive}: they update policies only from observed rewards and transitions, without learning a predictive model of the environment. world model–RL follows a different paradigm~\cite{wang2024meta}. It first learns a latent dynamics model that predicts future states and rewards, and then uses imagined rollouts inside this model to improve policies and value functions. Such designs, often built on recurrent state–space models, have shown good sample efficiency and long-horizon planning ability in continuous control and robotics~\cite{wu2023daydreamer}. Existing world model-RL frameworks, however, are mainly developed for generic control tasks~\cite{gao2024cardreamer,kessler2023effectiveness}. They are not tailored to MEC scenarios with coupled interference, multi-user QoS constraints, and non-convex latency structures. In particular, there is still a gap in designing a world model-PPO algorithm that remains on-policy and stable, while using imagination and internal dynamics modeling to perform foresightful offloading and power control in compact-LLM–based MEC systems.

\begin{figure}[t]
\centering{\includegraphics[width=0.42\textwidth]{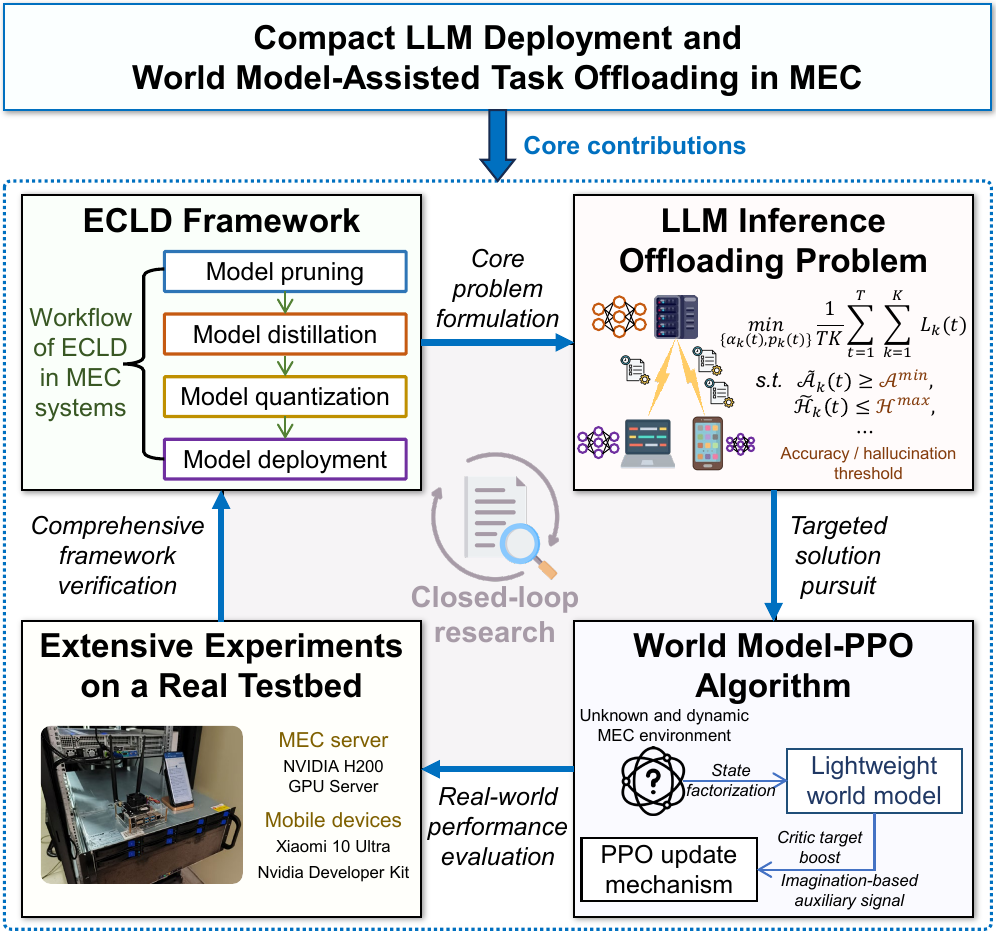}}
\caption{Overview of research contents in this paper, including the workflow of the proposed ECLD framework, the formulation of LLM inference offloading problem, the development of a world model-PPO algorithm for dynamic offloading, and comprehensive validation on a real MEC testbed.}\label{contributions}
\end{figure}

Motivated by these, this work investigates joint compact LLM deployment and MEC-based task offloading for next-generation edge intelligence. We consider a system where each mobile LLM user (MLU) runs a quantized compact LLM locally, while an edge server hosts a full-scale LLM to assist complex queries~\cite{zhang2025toward}. As shown in Fig.~\ref{contributions}, we first design an {Edge Compact LLM Deployment} (ECLD) framework that systematically applies pruning, distillation, and quantization to obtain device-specific compact models. We then formulate an optimization problem that jointly controls offloading ratios and transmit powers to minimize average inference latency under accuracy, hallucination, and energy constraints. Finally, we develop a {world model-PPO} algorithm that learns an adaptive offloading and power-control policy by combining on-policy PPO with a lightweight world model that captures MEC dynamics. \textit{To the best of our knowledge, this is the first work that couples a structured compact-LLM deployment pipeline with an LLM-aware MEC offloading formulation and a world-model–enhanced RL algorithm in a unified framework.} The main contributions are summarized as follows:

\begin{itemize}
  \item \textbf{Edge Compact LLM Deployment Framework Design:}
  We propose the ECLD framework that combines width and depth pruning, knowledge distillation, and low-bit quantization to obtain compact LLMs tailored to different edge devices. The framework defines importance scores for layers, neurons, heads, and embedding dimensions, derives a unified pruning mask, and applies hardware-aware quantization policies (e.g., 4-bit for smartphones and 8-bit for edge servers), forming a principled pipeline for on-device LLM deployment.

  \item \textbf{LLM-Aware MEC System Modeling and Problem Formulation:}
  We develop a compact-LLM MEC system where each MLU can flexibly split LLM inference between local execution and edge offloading. We introduce accuracy and hallucination as LLM-specific QoS metrics based on the alignment between local and edge-generated responses, and formulate a joint offloading and power-control problem that minimizes average inference latency subject to per-user energy budgets and global QoS constraints.

  \item \textbf{World model-PPO for Adaptive Offloading:}
  We design a world model-PPO algorithm to solve the above problem under unknown and time-varying MEC dynamics. A lightweight recurrent world model is trained to predict next states and rewards, and its outputs are used to construct mixed temporal-difference (TD) targets for the critic and an imagination-based auxiliary loss for the actor. This design improves sample efficiency and long-horizon decision making while preserving the on-policy stability of standard PPO.

  \item \textbf{Performance Evaluation and Insights:}
  We conduct extensive simulations to evaluate the proposed ECLD and world model-PPO framework against representative baselines. The results show that the deployment pipeline achieves notable reductions in model size and inference latency with limited accuracy loss, enabling practical operation on diverse edge devices. The control algorithm consistently reduces end-to-end latency while satisfying accuracy, hallucination, and energy constraints, and provides insights into how compression levels and offloading policies jointly shape the latency–energy–QoS trade-off.
\end{itemize}


\section{Related Work}
In this section, we review related studies on compact and edge-based LLM deployment, as well as world model–enhanced reinforcement learning for MEC, and highlight the key research gaps that motivate our work.

\subsection{Edge-Based LLM Deployment}
Edge-based LLM deployment refers to pushing LLM inference from centralized cloud data centers to edge servers that are geographically closer to end users, so that prompts are processed with reduced end-to-end latency and lower backbone traffic. Several recent studies have explored such deployments for generative LLMs. For example, Zhu \emph{et al.}~\cite{zhu2025birds} proposed {Birds in Cages}, an edge inference allocation framework that partitions an LLM across collaborating edge servers and optimizes the assignment of user requests to maximize overall service benefit under computation, storage, and latency constraints. Zhao \emph{et al.}~\cite{10693742} designed an edge–terminal cooperative scheme in which a small LLM on the terminal generates speculative tokens, while a large LLM on the edge verifies them in parallel; they jointly optimize model approximation and the number of speculative tokens via a branch-and-bound algorithm to minimize average delay and energy consumption. Lin \emph{et al.}~\cite{11152695} proposed a 6G MEC architecture for LLMs that leverages end–edge cooperation by combining split learning and inference, parameter-efficient fine-tuning, edge-side parallel training, and small–large model cooperation, thereby overcoming the latency, bandwidth, and privacy limitations of purely cloud- or device-based deployment.  Picano \textit{et al.}~\cite{10966456} modeled LLM layer placement as a two-sided matching game between transformer layers and heterogeneous edge nodes, jointly accounting for computation, communication latency, and bubble-time externalities to minimize end-to-end inference delay. Cai \textit{et al.}~\cite{10707514} proposed {Edge-LLM}, a server–node collaborative framework that splits the LLM backbone and adapter across edge servers and edge nodes and employs adaptive quantization, feature-map caching, and a value-density-first scheduler to accelerate inference and fine-tuning under resource constraints. Xu \textit{et al.}~\cite{xu2025decentralized} investigated decentralized LLM deployment over MEC networks and proposed a tensor-parallel parallel communication and computation (PCC) protocol that jointly optimizes model partitioning, task offloading, and result downloading across multiple edge servers to minimize inference latency. Muralidharan \textit{et al.}~\cite{muralidharan2024compact} developed {MINITRON}, which compressed a pretrained 15B Nemotron-4 LLM into 8B and 4B variants via structured depth/width pruning and data-efficient knowledge distillation, achieving $2$–$4\times$ parameter reduction while maintaining competitive downstream accuracy.

Despite this progress, most existing studies either focus on server-side layer placement and task allocation in distributed edge infrastructures, or on offline compression of a single LLM for a specific hardware target. They seldom provide a unified, \emph{hardware-aware} deployment pipeline that jointly coordinates pruning, distillation, quantization, and format conversion for diverse mobile and edge devices. Moreover, they do not explicitly couple the design of compact on-device LLMs with an MEC offloading policy. In particular, latency and energy are often optimized in isolation, while LLM-specific QoS aspects such as semantic accuracy and hallucination are not directly embedded into the deployment and resource-control decisions. These gaps motivate our joint design of a systematic compact LLM deployment framework, i.e., ECLD.

\subsection{World models for Wireless Communications}
World models have emerged as a promising tool in communication AI, since they allow agents to learn compact latent representations of network dynamics and to plan in imagination rather than relying purely on reactive interaction with the real environment~\cite{gao2024cardreamer,kessler2023effectiveness}. Several recent studies have begun to explore world-model-based designs in wireless and networking scenarios. For example, Wang \emph{et al.}~\cite{wang2025world} developed a world-model-based reinforcement learning framework for vehicular networks, where a learned latent dynamics model predicts the long-term evolution of the age of information (AoI) under time-varying channels and supports trajectory planning without extensive real-world exploration. Chai \emph{et al.}~\cite{chai2025mobiworld} proposed \emph{MobiWorld}, a generative world model that fuses heterogeneous and multimodal measurements to construct high-fidelity simulation environments for communication-aware network planning. Zhao \emph{et al.}~\cite{zhao2025world} introduced a world-model-driven agent framework for low-altitude economy networks and showed that latent imagination and predictive modeling can effectively handle partial observability and temporal dependencies in aerial communication systems, which improves decision-making for low-altitude communication and control. Most recently, Liu \emph{et al.}~\cite{liu2025dwm} proposed a decentralized world model with reasoning offloading (DWM-RO) framework, in which each base station-user pair is equipped with a recurrent state–space world model that learns compact predictive representations of SWIPT-enabled satellite–terrestrial networks dynamics and supports imagination-based multi-agent reinforcement learning for joint beamforming and power-splitting control.

Despite these advances, existing studies primarily employ world models for trajectory planning or high-fidelity offline network simulation and do not explicitly address joint offloading and power control in MEC systems equipped with compact LLMs and stringent QoS requirements. In particular, the integration of a lightweight world model with an on-policy, PPO algorithm for decision making under coupled latency, energy, accuracy, and hallucination constraints remains largely unexplored. This gap motivates the world model-PPO framework proposed in this work.


\begin{figure*}[t]
\centering{\includegraphics[width=0.80\textwidth]{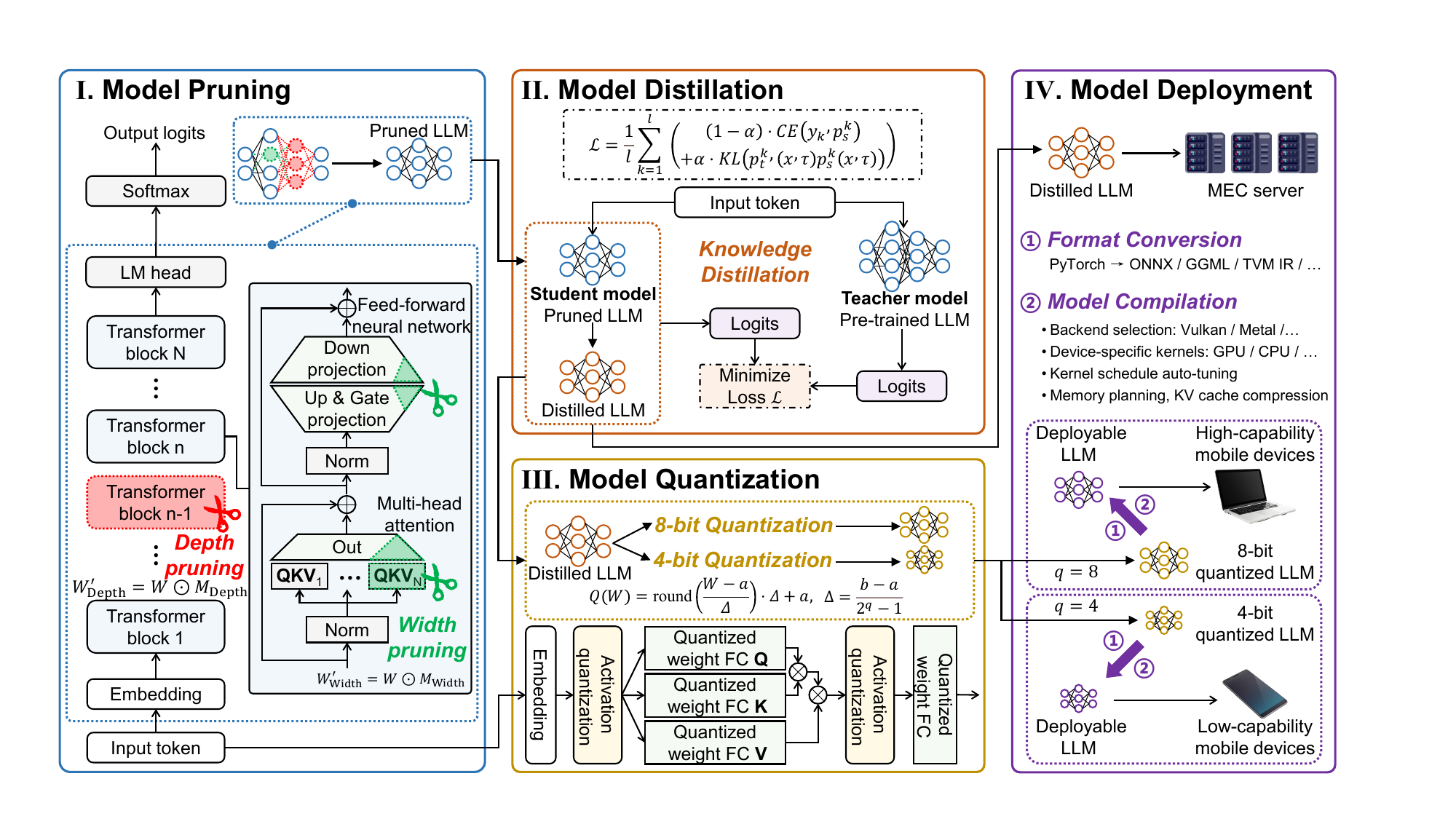}}
\caption{Workflow of the proposed ECLD framework for compact LLM deployment including four stages. \textbf{Stage \uppercase\expandafter{\romannumeral1}} is the sequential model pruning process for model size reduction. \textbf{Stage \uppercase\expandafter{\romannumeral2}} is the model distillation process through knowledge distillation for performance recovery. \textbf{Stage \uppercase\expandafter{\romannumeral3}} is the model quantization process for hardware efficiency. \textbf{Stage \uppercase\expandafter{\romannumeral4}} is the optimized model deployment process tailored to resource-constrained mobile and edge devices.}\label{Fig2}
\end{figure*}

\section{Edge Compact LLM Deployment}
Deploying LLMs on mobile and edge devices is challenging due to their massive parameter sizes and intensive computation and memory requirements, which often exceed the capabilities of resource-constrained platforms~\cite{zhang2024edgeshard}. To address these limitations, we develop an edge compact LLM deployment (ECLD) framework that progressively reduces the model size and complexity while preserving task performance. As illustrated in Fig.~\ref{Fig2}, ECLD consists of four key stages: (i) model pruning, (ii) model distillation via knowledge distillation, (iii) model quantization, and (iv) hardware-aware deployment. The detailed design of each stage is presented as follows.

\subsection{Model Pruning}
The first stage of the proposed ECLD framework is model pruning, which aims to reduce the size and computational complexity of an LLM by selectively removing parameters that have limited impact on its performance. We consider four types of structural components, namely layers, neurons, attention heads, and embedding dimensions, and assign each of them an importance score~\cite{zhao2025hape}. The corresponding importance scores, denoted by $S_{\text{Layer}}$, $S_{\text{Neuron}}$, $S_{\text{Head}}$, and $S_{\text{Embed}}$, capture the relative contribution of each component to the overall behavior of the LLM.

Formally, these importance scores are computed through an importance function defined as follows:
\begin{equation}
\{S_{\text{Layer}}, S_{\text{Neuron}}, S_{\text{Head}}, S_{\text{Embed}}\} = f(W, D),
\end{equation}
where $W$ represents the LLM parameters and $D$ denotes a small calibration dataset used for the evaluation. The function $f(\cdot)$ measures how sensitive the model predictions are to perturbations in the corresponding parameters. Components with higher sensitivity are assigned larger importance scores, whereas components with low scores are regarded as redundant or less critical and are therefore treated as pruning candidates~\cite{muralidharan2024compact}. In our framework, the pruning strategy is divided into two complementary components, i.e., width pruning and depth pruning, which are detailed as follows.

\subsubsection{\textbf{Width Pruning}}
Width pruning focuses on reducing the computational complexity of the LLM by removing neurons, attention heads, and embedding dimensions within each layer, while keeping the model depth unchanged. This procedure relies on the previously computed importance scores to identify components that contribute least to the model performance~\cite{Sharath2024llm}.

To describe this strategy, we introduce a binary pruning mask $M_{\text{Width}}$:
\begin{equation}
    M_{\text{Width}} = 
    \begin{cases}
        1, & \text{if } S_{\text{Neuron}}, S_{\text{Head}} \text{ or } S_{\text{Embed}} \geq \theta_{\text{pru}},\\[5pt]
        0, & \text{otherwise,}
    \end{cases}
\end{equation}
where $\theta_{\text{pru}}$ is a predefined pruning threshold. Components whose importance scores (i.e., $S_{\text{Neuron}}$, $S_{\text{Head}}$, or $S_{\text{Embed}}$) fall below this threshold are regarded as redundant and are therefore pruned.

The pruned weight tensor $W'_{\text{Width}}$ is obtained by applying this mask to the original weight tensor $W$:
\begin{equation}
    W'_{\text{Width}} = W \odot M_{\text{Width}},
\end{equation}
where $\odot$ denotes element-wise multiplication, i.e., Hadamard product. In this way, width pruning reduces the computational cost by removing non-critical components, while preserving the original depth structure of the LLM.

\subsubsection{\textbf{Depth Pruning}}
Depth pruning aims to reduce the number of layers in the LLM by removing entire layers or groups of layers that have limited influence on the overall model performance~\cite{muralidharan2024compact}. This layer-wise pruning is guided by the layer importance score $S_{\text{Layer}}$.

We define a binary depth pruning mask $M_{\text{Depth}}$ as
\begin{equation}
    M_{\text{Depth}} = 
    \begin{cases}
        1, & \text{if } S_{\text{Layer}} \geq \theta_{\text{pru}},\\[5pt]
        0, & \text{otherwise,}
    \end{cases}
\end{equation}
where $\theta_{\text{pru}}$ denotes the pruning threshold for layers. Layers with $S_{\text{Layer}}$ below this threshold are treated as less critical and are pruned.

The corresponding pruned weights are obtained by applying $M_{\text{Depth}}$ to the original weight tensor, i.e.,
\begin{equation}
    W'_{\text{Depth}} = W \odot M_{\text{Depth}}.
\end{equation}
By selectively eliminating non-essential layers, depth pruning reduces the model complexity and accelerates inference, while typically incurring only minor accuracy degradation.

\subsubsection{\textbf{Combined Pruning}}
To further compress the model, we combine width and depth pruning into a unified strategy that removes redundancy along both dimensions. We first construct a combined binary pruning mask $M$ by applying element-wise multiplication to the width and depth masks, i.e.,
\begin{equation}
    M = M_{\text{Width}} \odot M_{\text{Depth}},
\end{equation}
so that $M$ simultaneously encodes the pruning decisions for neurons, attention heads, embedding dimensions, and layers.

The final pruned weights $W'$ are then obtained by applying $M$ to the original weight tensor $W$, i.e.,
\begin{equation}
    W' = W \odot M.
\end{equation}
In this way, the combined pruning step produces a more compact and computationally efficient LLM, while retaining sufficient capacity for downstream tasks, which is essential for practical edge deployment.

\subsection{Model Distillation}

Although pruning effectively reduces redundancy and computational complexity, it inevitably leads to some performance degradation~\cite{muralidharan2024compact}. To mitigate this effect, we include a knowledge distillation stage in the ECLD framework, where the original full-scale LLM (i.e., the {teacher} model) transfers its knowledge to the pruned LLM (i.e., the {student} model). The goal is for the student to approximate the teacher’s behavior while preserving a compact and computation-efficient structure~\cite{yang2025survey}.

Formally, the distillation process aligns the output distributions of the student and teacher by minimizing a joint loss, which is given by
\begin{equation}
    \mathcal{L} = \frac{1}{l} \sum_{k=1}^{l} \Big[ (1 - \alpha)\,\mathrm{CE}(y_k, p_s^k) 
    + \alpha\,\mathrm{KL}\big(p_t^k(x, \tau), p_s^k(x, \tau)\big) \Big],
\end{equation}
where the cross-entropy (CE) loss aligns the student outputs with the ground-truth labels, i.e.,
\begin{equation}
    \mathrm{CE}(y_k, p_s^k) 
    = - \sum_{i=1}^{I} y_{k,i} \log(p_{s,i}^k),
\end{equation}
and the Kullback–Leibler (KL) divergence aligns the softened predictions of the teacher and student, i.e.,
\begin{equation}
    \mathrm{KL}\big(p_t^k(x, \tau), p_s^k(x, \tau)\big)
    = \sum_{i=1}^{I} p_{t,i}^k(x, \tau) 
      \log\frac{p_{t,i}^k(x, \tau)}{p_{s,i}^k(x, \tau)}.
\end{equation}
Here, $\alpha$ is a balancing hyperparameter that controls the trade-off between fitting hard labels (i.e., ground truth) and soft labels (i.e., teacher outputs). The softmax temperature $\tau$ is used to smooth the teacher’s output distribution, expressed as
\begin{equation}
    O(x, \tau) = \mathrm{softmax}\!\left(\frac{z(x)}{\tau}\right),
\end{equation}
where $z(x)$ denotes the logits produced by the teacher model.

\begin{remark}
The softmax temperature $\tau$ facilitates knowledge distillation by smoothing the teacher’s output distribution. Larger values of $\tau$ lead to a more uniform probability profile, which reveals fine-grained relationships among classes and helps the student learn nuanced decision boundaries from the teacher.
\end{remark}

By jointly aligning the student with both hard labels and soft teacher-generated probabilities, the distillation stage can recover much of the performance loss caused by pruning, allowing the pruned model to retain high accuracy despite its reduced complexity~\cite{Sharath2024llm}.

\subsection{Model Quantization}

Although pruning and distillation substantially reduce the size and complexity of LLMs, the resulting models can still be demanding for resource-limited edge devices~\cite{shen2024agile}. As the third stage of the ECLD framework, we therefore apply model quantization, which converts high-precision floating-point weights into lower-precision numerical representations. This further step cuts both memory usage and computation cost and makes the compact LLM more suitable for mobile and edge platforms~\cite{10596048}.

Concretely, the quantization process maps the original floating-point weight tensor $W$ to discrete levels via
\begin{equation}
    Q(W) = \text{round}\!\left(\frac{W - a}{\Delta}\right) \cdot \Delta + a,
\end{equation}
where the quantization step size $\Delta$ is given by
\begin{equation}
    \Delta = \frac{b - a}{2^q - 1}.
\end{equation}
Here, $q$ denotes the bit-width used for quantization (e.g., 4-bit or 8-bit), and $a$ and $b$ are the minimum and maximum values of $W$, respectively. The operator $\text{round}(\cdot)$ maps each weight to its nearest quantization level, which allows inference to be implemented with efficient integer operations.

With this quantization stage, the ECLD framework achieves additional reductions in memory footprint and computational load while preserving sufficient numerical precision for accurate inference. To better match different hardware platforms, we further adopt a hardware-aware quantization strategy as follows~\cite{juneja2025halo}.

\subsubsection{\textbf{Hardware-Aware Quantization}}

To ensure both compatibility and efficiency across different hardware platforms, we select the quantization bit-width according to the computational and memory constraints of the target device:
\begin{itemize}
    \item \textbf{4-bit quantization:} Provides the largest memory savings and lowest energy consumption, and is therefore suitable for smartphones and very resource-limited mobile devices.
    \item \textbf{8-bit quantization:} Achieves a good balance between computational efficiency and model accuracy, and is adopted for laptops and edge servers with relatively stronger resources.
\end{itemize}
In this way, the bit-width is chosen in a hardware-aware manner so that both resource utilization and overall system performance are jointly improved~\cite{juneja2025halo}.

\subsubsection{\textbf{Quantization Error Minimization}}

To limit the performance loss introduced by quantization, we further refine the quantization parameters through the following optimization problem, i.e., $\min_{a, b} \, \| W - Q(W) \|_2^2 
    = \min_{a, b} \sum_{i} \big(W_i - Q(W_i)\big)^2$, where $\|\cdot\|_2$ denotes the $\ell_2$ norm, $W$ is the original weight tensor, and $Q(W)$ is its quantized counterpart. By choosing $(a,b)$ that best align the quantized weights with their high-precision values, the quantization error can be effectively controlled~\cite{10596048}.

\subsection{Compact LLM Deployment}

By integrating pruning, distillation, and quantization, the original LLM is transformed into a lightweight and computationally efficient model that can be deployed on a wide range of edge and mobile devices. As illustrated in Stage~IV of Fig.~\ref{Fig2}, the quantized LLM is then adapted to the target hardware through format conversion and model compilation. In this stage, the model is compiled into optimized binaries, integrated with hardware-specific runtimes and tokenizers, and configured according to device capabilities (e.g., 4-bit or 8-bit execution)~\cite{tvm}. The final deployment is validated using key metrics such as task accuracy, energy consumption, and inference latency, so as to ensure reliable operation under realistic resource constraints.

\begin{example}
For the Llama-3 8B model, a 4-bit quantized compact variant can be deployed on a smartphone with more than 6~GB of RAM and can reach an inference speed of approximately 2--3 tokens per second~\cite{zhang2025toward}. On a laptop equipped with a GPU with at least 8~GB of VRAM (e.g., NVIDIA RTX~3060), an 8-bit quantized compact variant of the same model can be deployed, delivering an inference speed of around 8--10 tokens per second.
\end{example}

After applying ECLD, each mobile user is endowed with a device-specific compact LLM whose latency, energy consumption, and quality of generation can be explicitly quantified. Specifically, the role of ECLD is to turn a monolithic cloud-scale model into a family of hardware-aware local models with known performance–cost trade-offs. Once these local profiles are fixed, the remaining degree of freedom lies in {how} each inference task is split between local execution and edge assistance over time.

\section{System Model and Problem Formulation}

With compact LLMs deployed on multiple mobile edge users through the proposed ECLD framework, practical limitations still arise due to the inherently constrained computation and energy resources of user devices. Consequently, a systematic task offloading mechanism is required to efficiently coordinate and allocate LLM inference workloads between local devices and the edge server~\cite{10591707}. In what follows, we model this network-level decision problem by explicitly incorporating the latency, energy, and QoS characteristics of the compact LLMs obtained from ECLD.

We consider an MEC system comprising a single edge server that hosts a high-capacity, full-scale LLM and a set of $K$ mobile LLM users (MLUs), each equipped with a locally deployed quantized compact LLM. The set of MLUs is denoted by $\mathcal{K} = \{1,\dots,K\}$. Time is slotted and indexed by $t \in \mathcal{T} = \{1,\dots,T\}$, where each slot has duration $\tau$ seconds.

At the beginning of each time slot $t \in \mathcal{T}$, every MLU $k \in \mathcal{K}$ receives an LLM inference task (e.g., semantic parsing, intent recognition, or instruction following), characterized by an input of size $X_k(t)$ bits. To accommodate heterogeneous task complexities and device capabilities, each task can be partitioned: a fraction is processed locally by the compact LLM on the MLU, while the remaining fraction is offloaded to the edge server running the full-scale LLM~\cite{10591707}. This cooperative processing paradigm improves overall accuracy and mitigates hallucination effects that may arise from purely local inference on heavily compressed models~\cite{zhang2025toward}.





\begin{figure}[t]
\centering{\includegraphics[width=0.42\textwidth]{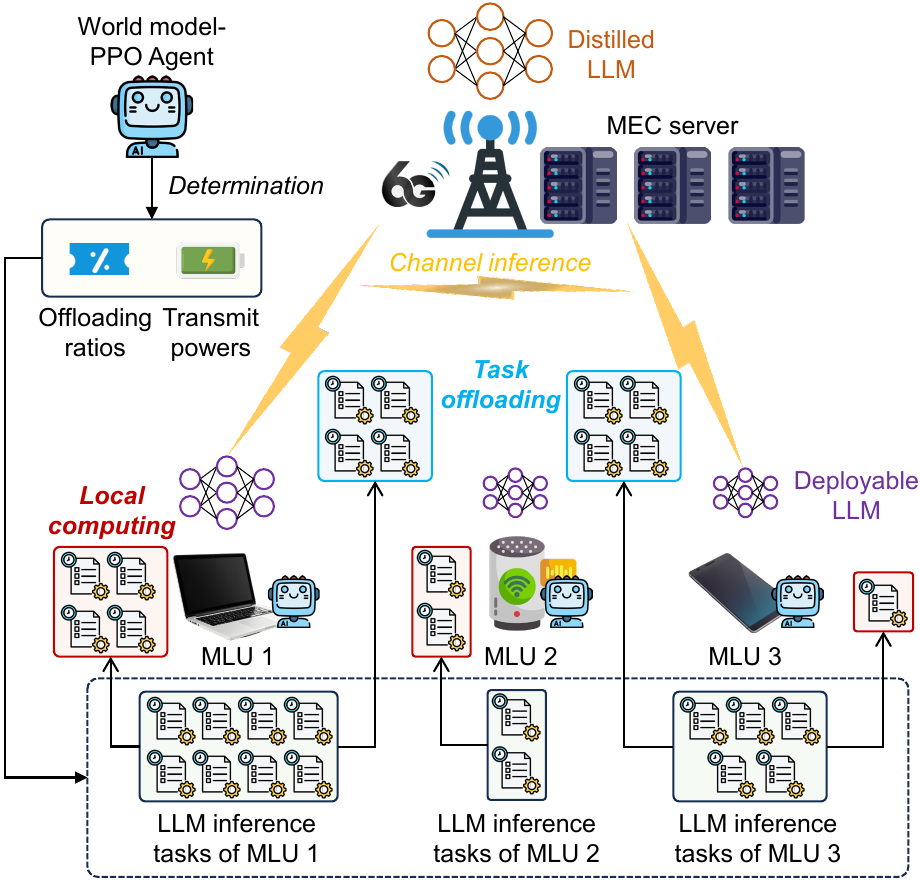}}
\caption{System model of cooperative LLM inference. Each mobile LLM user partitions its task between local execution using a compact quantized LLM and remote execution via uplink offloading to an MEC server hosting a distilled LLM.}\label{system_model}
\end{figure}

\subsection{Local Computing Model}

Let $\alpha_k(t)\in[0,1]$ denote the offloading ratio of MLU $k$ at time slot $t$, so that a fraction $(1-\alpha_k(t))X_k(t)$ bits of the input is processed locally. Let $f_k$ be the CPU clock frequency of MLU $k$ (in cycles per second), and let $\phi$ denote the number of CPU cycles required per bit for inference with the quantized compact LLM. The local computation latency of MLU $k$ at time slot $t$ is then given by
\begin{equation}
    L_k^{\text{local}}(t) 
    = \frac{(1-\alpha_k(t))\,\phi X_k(t)}{f_k}.
\end{equation}

The corresponding local computation energy consumption is given by
\begin{equation}
    E_k^{\text{local}}(t) 
    = \kappa_{\text{comp}} f_k^2 (1-\alpha_k(t))\,\phi X_k(t),
\end{equation}
where $\kappa_{\text{comp}}$ is a device-specific computation energy coefficient. Because the local model is pruned and quantized, purely on-device inference may suffer from reduced accuracy and a higher risk of hallucination~\cite{zhang2025toward}. The impact of local processing on accuracy and hallucination is explicitly characterized in Section~\ref{subsec:accuracy}.

\subsection{Channel Model}

We consider a two-dimensional deployment of the MLUs and the MEC server~\cite{11152695}. The position of MLU $k$ is denoted by $\big[q_k^x, q_k^y\big]$, and the MEC server is located at horizontal coordinates $\big[q_{\text{MEC}}^x, q_{\text{MEC}}^y\big]$ with antenna height $h_{\text{MEC}}$. The Euclidean distance between MLU $k$ and the MEC server is given by
\begin{equation}
    d_k = \sqrt{\big(q_k^x - q_{\text{MEC}}^x\big)^2 
    + \big(q_k^y - q_{\text{MEC}}^y\big)^2 
    + h_{\text{MEC}}^2}.
\end{equation}

The wireless channel gain between MLU $k$ and the MEC server at time slot $t$ consists of large-scale path loss and small-scale fading, and is modeled as
\begin{equation}
    h_k(t) = g_k \tilde{h}_k(t),
\end{equation}
where $g_k$ captures the large-scale path loss and follows a distance-dependent model, i.e., $g_k = \frac{g_0}{d_k^2}$, with $g_0$ denoting the reference channel gain at a distance of $1$\,m. The small-scale fading term $\tilde{h}_k(t)$ is modeled as Rician fading, which is given by
\begin{equation}
    \tilde{h}_k(t) = \left|\sqrt{\frac{\kappa}{\kappa + 1}} 
    + \sqrt{\frac{1}{\kappa + 1}}\,\bar{h}_k(t)\right|^2,
\end{equation}
where $\bar{h}_k(t)\sim\mathcal{CN}(0,1)$ is a circularly symmetric complex Gaussian random variable, and $\kappa$ is the Rician $K$-factor that characterizes the power ratio between the line-of-sight and scattered components.

\subsection{Offloading Model}

Due to the limited uplink bandwidth, simultaneous transmissions from multiple MLUs cause mutual interference~\cite{10591707}. The uplink achievable data rate between MLU $k$ and the MEC server at time slot $t$ is modeled as
\begin{equation}
    r_k(t) = B \log_2\!\left(1 + 
    \frac{p_k(t) h_k(t)}{\sum_{l\in\mathcal{K},\, l\neq k} p_l(t) h_l(t) + \sigma^2}\right),
\end{equation}
where $B$ denotes the uplink bandwidth, $p_k(t)$ is the transmit power of MLU $k$, and $\sigma^2$ denotes the noise power.

Given $r_k(t)$, the uplink offloading latency of MLU $k$ at time slot $t$ is modeled as
\begin{equation}
    L_k^{\text{off}}(t) = \frac{\alpha_k(t) X_k(t)}{r_k(t)},
\end{equation}
and the corresponding uplink transmission energy consumption is modeled as
\begin{equation}
    E_k^{\text{off}}(t) = p_k(t)\,L_k^{\text{off}}(t).
\end{equation}

The MEC server is equipped with a total computational capability $F$ (in CPU cycles per second). The computation latency for processing the offloaded portion of MLU $k$'s task is modeled as
\begin{equation}
    L_k^{\text{MEC}}(t) = \frac{\alpha_k(t)\,\phi X_k(t)}{F},
\end{equation}
where $\phi$ is the number of CPU cycles required per bit for LLM inference, consistent with the local computing model.

\subsection{Online System-Level QoS Metrics}\label{subsec:accuracy}

Conventional MEC formulations typically focus on latency and energy minimization. In LLM-driven applications, however, the {quality} of generated content is equally critical. A solution that is fast but inaccurate, or that frequently hallucinates, is unacceptable in many practical scenarios such as question answering and decision support. Therefore, in addition to latency and energy, we explicitly characterize two QoS indicators, i.e., {accuracy} and {hallucination rate}. These indicators capture complementary aspects of LLM reliability and will later be enforced as constraints in our optimization problem to ensure acceptable generation quality.

\subsubsection{Offline Model-Level Evaluation}

Before deployment, we benchmark each LLM using standard datasets to obtain aggregate measures of accuracy and hallucination. Let \(\bar{\mathcal{A}}\) and \(\bar{\mathcal{H}}\) denote the offline accuracy and hallucination metrics, respectively.

\textbf{Accuracy:} Accuracy measures the proportion of correct responses generated by the LLM on a natural language question--answer (QA) dataset containing \(N_a\) QA pairs~\cite{yih2016value}. A response \(\hat{y}_{n_a}\) is counted as correct if it contains the ground-truth answer \(y_{n_a}\) anywhere in the generated text. Let \(\mathbb{I}[\hat{y}_{n_a} \supseteq y_{n_a}]\) be an indicator function that equals \(1\) if \(\hat{y}_{n_a}\) contains \(y_{n_a}\), and \(0\) otherwise. The offline accuracy is then given by
\begin{equation}
    \bar{\mathcal{A}} 
    = \frac{1}{N_a}\sum_{n_a=1}^{N_a} 
    \mathbb{I}[\hat{y}_{n_a} \supseteq y_{n_a}].
    \label{eq:acc}
\end{equation}

\textbf{Hallucination:} Hallucination behavior is evaluated using a dataset composed of Wikipedia articles \(R_i\) (e.g., biographies of notable individuals)~\cite{manakul2023selfcheckgpt}. The model is given only the initial portion of each article (such as the title and opening sentences) and is instructed to complete the remaining content. Let the LLM output for article \(i\) be \(\hat{y}_i = \{s_{i,1},\dots,s_{i,S_i}\}\), where \(s_{i,j}\) is the \(j\)-th generated sentence and \(S_i\) is the number of generated sentences. We define an indicator function \(\mathbb{I}_{\text{fact}}(s_{i,j},R_i)\) that equals \(1\) if sentence \(s_{i,j}\) is factually consistent with the ground-truth article \(R_i\), and \(0\) otherwise. The offline hallucination rate is then given by
\begin{equation}
    \bar{\mathcal{H}} 
    = 1 - \frac{\sum_{i=1}^{N_h}\sum_{j=1}^{S_i} \mathbb{I}_{\text{fact}}(s_{i,j},R_i)}
    {\sum_{i=1}^{N_h} S_i},
    \label{eq:hallu}
\end{equation}
where \(N_h\) is the number of articles in the hallucination evaluation set. A smaller value of \(\bar{\mathcal{H}}\) indicates fewer hallucinations and higher factual reliability. These offline metrics are used to compare different compression strategies (pruning, distillation, quantization) and to select candidate compact LLMs for deployment in the MEC system.

Since the offloading ratio \(\alpha_k(t)\) determines how each task is split between local processing and MEC offloading, the effective slot-wise accuracy and hallucination perceived by MLU \(k\) are given by
\begin{equation}
\left\{
\begin{aligned}
    \tilde{\mathcal{A}}_k(t) &= \alpha_k(t)\cdot \mathcal{A}_k^\text{MEC}(t) 
    + \big(1-\alpha_k(t)\big)\mathcal{A}_k(t), \\
    \tilde{\mathcal{H}}_k(t) &= \alpha_k(t)\cdot \mathcal{H}_k^\text{MEC}(t)+ \big(1-\alpha_k(t)\big)\mathcal{H}_k(t),
\end{aligned}
\right.
\end{equation}
In the subsequent problem formulation, these QoS indicators will be enforced as constraints by requiring \(\tilde{\mathcal{A}}_k(t)\) to exceed a minimum accuracy threshold and \(\tilde{\mathcal{H}}_k(t)\) to stay below a maximum hallucination level, while latency and energy are optimized jointly~\cite{zhang2025toward}.

\subsection{Problem Formulation}

Our objective is to minimize the average inference latency experienced by all MLUs over the considered time horizon, while satisfying constraints on transmit power, per-slot energy consumption, and generation quality (in terms of accuracy and hallucination rate)~\cite{chen2021drl,lee2024exploring}. The overall latency of MLU $k$ at time slot $t$ is defined as
\begin{equation}\label{time_latency}
    L_k(t) = \max\big\{L_k^{\text{local}}(t),\; L_k^{\text{off}}(t)+L_k^{\text{MEC}}(t)\big\},
\end{equation}
where $L_k^{\text{local}}(t)$ is the local computation latency, $L_k^{\text{off}}(t)$ is the uplink transmission latency, and $L_k^{\text{MEC}}(t)$ is the computation latency at the MEC server.

At each time slot $t$, the system optimizes two decision variables for each MLU $k$: the offloading ratio $\alpha_k(t)$, which determines how the task is partitioned between local and MEC processing, and the transmit power $p_k(t)$ used for offloading. The joint optimization problem is formulated as
\begin{subequations}\label{P1}
    \begin{align}
        \min_{\{\alpha_k(t),\,p_k(t)\}}\quad 
        &\frac{1}{TK}\sum_{t=1}^{T}\sum_{k=1}^{K}L_k(t)\label{obj}\\[4pt]
        \text{s.t.}\quad
        &0 \le p_k(t)\le P_k^{\max},\label{cons_power}\\[4pt]
        &0 \le \alpha_k(t)\le 1,\label{cons_alpha}\\[4pt]
        &E_k^{\text{local}}(t)+E_k^{\text{off}}(t)\le E_k^{\max},\label{cons_energy}\\[4pt]
        &\tilde{\mathcal{A}}_k(t)\ge\mathcal{A}^{\min},\label{cons_acc}\\[4pt]
        &\tilde{\mathcal{H}}_k(t)\le\mathcal{H}^{\max},\label{cons_hallu}\\[4pt]
        &\forall k\in\mathcal{K},\; t\in\mathcal{T},\label{cons_indices}
    \end{align}
\end{subequations}
where $P_k^{\max}$ denotes the maximum transmit power of MLU $k$, and $E_k^{\max}$ is its per-slot energy budget. The QoS thresholds $\mathcal{A}^{\min}$ and $\mathcal{H}^{\max}$ represent the required minimum effective accuracy and maximum tolerable hallucination rate, respectively. 

Constraint~\eqref{cons_power} restricts the transmit power of each MLU to a feasible range and prevents excessive interference or hardware saturation. Constraint~\eqref{cons_alpha} specifies the admissible range of the offloading ratio and thus governs the fraction of each task executed locally versus offloaded to the MEC server. Constraint~\eqref{cons_energy} ensures that the total per-slot energy consumption of MLU $k$, including both local computation and uplink transmission, does not exceed its energy budget $E_k^{\max}$. Constraints~\eqref{cons_acc} and~\eqref{cons_hallu} enforce QoS requirements on generation accuracy and hallucination, and hence avoid offloading and power-control policies that are fast but unreliable.

Problem~\eqref{P1} is a stochastic and strongly coupled optimization problem. The objective and constraints are non-convex due to the interference-limited rate expression in the offloading model and the max-operator in~\eqref{time_latency}. In addition, the system dynamics, including time-varying channels, task arrivals, and LLM response quality, are only partially known in advance~\cite{10591707}. Therefore, we propose a learning-based solution, i.e., world model-PPO algorithm, to address these challenges in a sample-efficient and stable manner.

\section{World model-PPO Solution}

\begin{figure*}[t]
\centering{\includegraphics[width=0.80\textwidth]{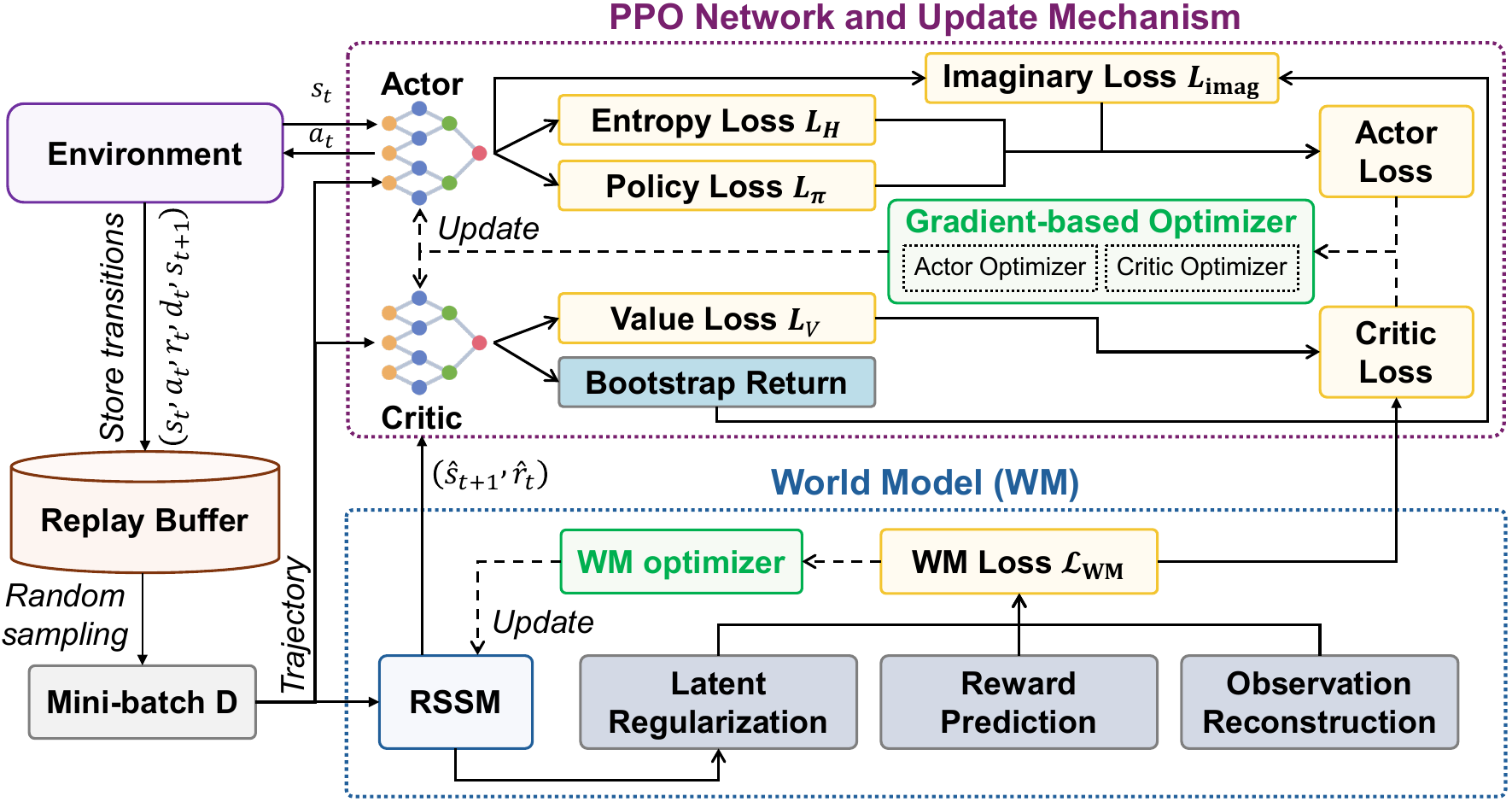}}
\caption{Architecture of the world model-PPO, where the actor–critic networks of PPO are jointly updated with a lightweight RSSM-based world model that predicts latent dynamics, reconstructs observations, and supports imagination-based policy improvement.}\label{PPO_algorithm}
\end{figure*}

In this section, we develop a learning-based control framework that adaptively determines offloading and power decisions for the compact-LLM MEC system. At its core, we adopt an on-policy PPO agent that repeatedly observes the environment, evaluates long-term performance, and selects actions~\cite{chen2021drl}. On top of PPO, we introduce a lightweight world model that captures the evolution of task arrivals, wireless channels, and QoS indicators, and that augments PPO with model-based value estimation and imagination~\cite{wang2025world}. The resulting world model-PPO algorithm leverages both real experience and learned environment dynamics while preserving the stability of standard PPO, as illustrated in Fig.~\ref{PPO_algorithm}.

\subsection{PPO for Compact-LLM MEC}

We model the compact-LLM MEC control task as a Markov decision process (MDP) with state $s_t$, action $a_t$, and reward $r_t$. PPO is an on-policy actor–critic method that maintains a stochastic policy (i.e., actor) $\pi_\theta(a_t|s_t)$ with parameters $\theta$ and a value function (i.e., critic) $V_\phi(s_t)$ with parameters $\phi$. Given a batch of on-policy trajectories, PPO maximizes a clipped surrogate objective that constrains the policy update within a trust region. Let
\begin{equation}
    \rho_t(\theta) = \frac{\pi_\theta(a_t|s_t)}{\pi_{\theta_{\text{old}}}(a_t|s_t)}
\end{equation}
denote the probability ratio between the updated and behavior policies, and let $A_t$ be an estimate of the advantage function (e.g., obtained via generalized advantage estimation). The clipped PPO objective is given by
\begin{align}
\mathcal{L}_{\text{CLIP}}(\theta)
&= \mathbb{E}_t \Big[
\min\big( \rho_t(\theta) A_t,\;
\mathrm{clip}\big(\rho_t(\theta), 1-\epsilon, 1+\epsilon\big)\, A_t
\big)
\Big],
\end{align}
where $\epsilon>0$ is a hyperparameter that controls the size of the trust region. The critic is trained to regress the value function toward a target $V'_t$ (e.g., an $n$-step or $\lambda$-return) via the mean-squared error loss, which is given by
\begin{equation}
\mathcal{L}_{\rm V}(\phi) = \mathbb{E}_t \big[
\big(V_{\phi}(s_t)-V'_t\big)^2
\big].
\end{equation}

In the considered MEC setting, PPO iteratively updates the policy by using the observed latency, energy consumption, and QoS feedback~\cite{10591707}. At each time slot, the agent observes the state $s_t$, samples the offloading and power decisions $a_t$ from $\pi_\theta(a_t|s_t)$, and then receives the resulting reward $r_t$ and next state $s_{t+1}$. By repeating this interaction, the actor and critic are jointly optimized, and the learned policy adapts to non-stationary traffic, channel conditions, and user-level QoS requirements in an online manner.

\subsection{World model }

While PPO can learn a suitable policy from direct interaction with the environment, pure trial-and-error in a high-dimensional and stochastic MEC system may require a large number of samples. To improve data efficiency and provide limited foresight, we introduce a world model, which is a learned latent dynamics model that predicts how the environment evolves in response to actions~\cite{wang2025world}. The world model is lightweight compared with the underlying LLMs and is trained only on compact state–action–reward–observation trajectories.

We adopt a recurrent state-space model (RSSM) that factorizes the latent state into a deterministic recurrent component $h_t$ and a stochastic latent variable $z_t$. Given the previous latent state and action, the RSSM updates as
\begin{equation}
\left\{
\begin{aligned}
    h_t &= f_{\theta}(h_{t-1}, z_{t-1}, a_{t-1}),\\
    z_t &\sim q_{\phi}(z_t \mid h_t, o_t),
\end{aligned}
\right.
\end{equation}
where $o_t$ denotes the observed environment features (such as task size, channel state, and QoS feedback), $f_{\theta}$ is a recurrent transition model, and $q_{\phi}$ is an encoder that infers the stochastic latent from the current observation and recurrent state. A learned prior $p_{\theta}(z_t|h_t)$ and a set of decoders map the latent state $(h_t,z_t)$ to predicted observations $\hat{o}_t$ and rewards $\hat{r}_t$.

The world model is trained by minimizing a multi-term loss that balances reconstruction, reward prediction, and latent regularization~\cite{chai2025mobiworld}, which is given by
\begin{equation}
\begin{aligned}
\mathcal{L}_{\rm WM} 
    &= \underbrace{\mathbb{E}\big[\| \hat{o}_t - o_t \|_2^2\big]}_{\text{observation reconstruction}}
    + \underbrace{\lambda_r\,(\hat{r}_t - r_t)^2}_{\text{reward prediction}} \\
    &\quad
    + \underbrace{\beta\,\mathrm{KL}\big(q_{\phi}(z_t|h_t,o_t)\,\big\|\,p_{\theta}(z_t|h_t)\big)}_{\text{latent regularization}},
\end{aligned}
\end{equation}
where $\lambda_r$ and $\beta$ are weighting coefficients and $\mathrm{KL}(\cdot\|\cdot)$ denotes the Kullback–Leibler divergence.

After training, the world model can be rolled forward from a given state by propagating $(h_t,z_t)$ under hypothetical action sequences. This generates imagined future observations and rewards without additional interaction with the real MEC environment and allows PPO to exploit internal simulations when updating its policy.

\subsection{World model-PPO}

We now integrate the world model with PPO to use both real and imagined experience while preserving the stability of on-policy updates. The integration follows two principles: (i) using the world model to {boost} the critic targets, and (ii) providing a small imagination-based auxiliary signal for the actor.

Given real transitions $(s_t,a_t,r_t,d_t,s_{t+1})$, where $d_t$ is the episode termination flag, the world model predicts a one-step next state and reward $(\hat{s}_{t+1},\hat{r}_t)$ from $(s_t,a_t)$. We then form a mixed target for the value function by blending real and model-based TD targets, which is given by
\begin{equation}
\label{eq:boosted-target}
\begin{aligned}
y_t ={}&(1-\mu_{\text{wm}})\Big[r_t + \gamma (1-d_t)\,V_\phi(s_{t+1})\Big] \\
&\quad+ \mu_{\text{wm}}\Big[\hat{r}_t + \gamma (1-d_t)\,V_\phi(\hat{s}_{t+1})\Big],
\end{aligned}
\end{equation}
where $\gamma$ is the discount factor and $\mu_{\text{wm}}\in[0,1]$ controls the contribution of the world model. The critic is trained with
\begin{equation}
\label{eq:crit-wm}
\mathcal{L}_V = \mathbb{E}_t\big[(V_\phi(s_t) - y_t)^2\big],
\end{equation}
and the corresponding advantages are computed as $A_t = y_t - V_\phi(s_t)$ (optionally normalized). This value boosting reduces variance and allows the critic to exploit the predictive power of the world model while remaining anchored to real environment feedback~\cite{wang2024meta}.

The actor still uses the clipped PPO surrogate, now expressed with the boosted advantages, i.e.,
\begin{equation}
\label{eq:ppo}
\mathcal{L}_{\pi} = -\,\mathbb{E}_t\!\left[
    \min\Big(\rho_t(\theta) A_t,\;
             \mathrm{clip}\big(\rho_t(\theta),1-\epsilon,1+\epsilon\big)\,A_t\Big)
   \right],
\end{equation}
where
\begin{equation}
\rho_t(\theta) = \frac{\pi_\theta(a_t|s_t)}{\pi_{\theta_{\text{old}}}(a_t|s_t)}
\end{equation}
denotes the probability ratio between the updated and behavior policies, and $\epsilon>0$ controls the trust-region size.

To exploit imagination only where the world model is reliable, we first select a subset of low-uncertainty states. From each such state $S_0$, we roll out $H$ imagined steps using the world model, obtaining imagined trajectories $\{(S_t,A_t,R_t)\}_{t=0}^{H-1}$ and a terminal state $S_H$. Based on this imagined rollout, we compute a bootstrapped return as
\begin{equation}
G = \sum_{t=0}^{H-1}\gamma^t R_t + \gamma^H V_\phi(S_H).
\end{equation}
We then define an auxiliary imagination loss as
\begin{equation}
\label{eq:imag}
\mathcal{L}_{\text{imag}} = 
\eta\,\frac{1}{H}\sum_{t=0}^{H-1}
\mathbb{E}\!\big[\,-\,\underbrace{\big(G - V_\phi(S_t)\big)}_{\text{stop-gradient}}\,
\log \pi_\theta(A_t|S_t)\big],
\end{equation}
where $\eta>0$ scales the contribution of imagination and the term $\big(G - V_\phi(S_t)\big)$ is treated as a constant (no gradient) to avoid destabilizing the critic. This loss nudges the policy toward actions that achieve high long-horizon returns under the world model.

To encourage exploration and prevent premature convergence, we also add an entropy regularization term, i.e.,
\begin{equation}
\mathcal{L}_H = -\beta_{\text{ent}}\,\mathbb{E}_t\big[-\log \pi_\theta(a_t|s_t)\big],
\end{equation}
with entropy coefficient $\beta_{\text{ent}}>0$. The overall objectives for the actor and critic are respectively given by
\begin{equation}
\label{eq:overall-actor}
\mathcal{L}_{\text{actor}} = \mathcal{L}_{\pi} + \mathcal{L}_{\text{imag}} + \mathcal{L}_H,
\end{equation}
and
\begin{equation}
\label{eq:overall-critic}
\mathcal{L}_{\text{critic}} = \mathcal{L}_V + \lambda_{\text{wm}}\,\mathcal{L}_{\rm WM},
\end{equation}
where $\lambda_{\text{wm}}$ balances the world model training loss and critic fitting when the world model is jointly updated.  Based on this design, the world model-PPO algorithm maintains the robustness and on-policy stability of standard PPO, while enabling the agent to react to current MEC states, anticipate future latency and QoS through the learned world model, and improve offloading and power decisions using both real and imagined experience~\cite{kessler2023effectiveness}.

\begin{algorithm}[t]
    \caption{World model-PPO for Compact LLM Offloading}
    \label{alg:wm_ppo}
    \textbf{Input:} Training iterations $N_{\text{iter}}$, episode length $T$, discount factor $\gamma$, PPO clip parameter $\epsilon$, world model weight $\mu_{\rm wm}$, etc.\\
    \textbf{Output:} Policy $\pi_\theta$ for offloading and power control.\\[2pt]
    \For{$i=1$ \KwTo $N_{\text{iter}}$}{
        \textbf{(1) Trajectory collection}\\
        Reset environment and observe $s_0$;\\
        \For{$t=0$ \KwTo $T-1$}{
            Sample $a_t=\{\alpha_k(t),p_k(t)\}_{k\in\mathcal{K}} \sim \pi_\theta(\cdot|s_t)$;\\
            Execute $a_t$, observe $s_{t+1}$ and termination flag $d_t$;\\
            Compute reward $r_t$ from latency and penalty term $\Omega(t)$;\\
            Store $(s_t,a_t,r_t,d_t,s_{t+1})$ in $\mathcal{D}$;\\
            \If{$d_t=1$}{\textbf{break}}
        }
        \textbf{(2) World-model update}\\
        Sample mini-batches from $\mathcal{D}$ and update $\psi$ by minimizing $\mathcal{L}_{\rm WM}$;\\
        \textbf{(3) Critic update (world model-boosted targets)}\\
        Sample mini-batches from $\mathcal{D}$; predict $(\hat{s}_{t+1},\hat{r}_t)$ with the world model;\\
        Form mixed targets $y_t$ via \eqref{eq:boosted-target} and advantages $A_t = y_t - V_\phi(s_t)$;\\
        Update $\phi$ by minimizing $\mathcal{L}_V = \mathbb{E}[(V_\phi(s_t)-y_t)^2]$;\\
        \textbf{(4) Actor update (PPO + imagination)}\\
        Compute PPO loss $\mathcal{L}_\pi$ with $A_t$;\\
        From selected states, roll out $H$ world model steps, compute imagined return $G$ and loss $\mathcal{L}_{\text{imag}}$;\\
        Add entropy loss $\mathcal{L}_H$ and update $\theta$ by descending
        $\mathcal{L}_{\text{actor}} = \mathcal{L}_\pi + \mathcal{L}_{\text{imag}} + \mathcal{L}_H$;
    }
\end{algorithm}

\subsection{MDP Formulation}

We formulate the compact-LLM MEC control problem as an MDP. A centralized controller deployed at the MEC server acts as the \emph{agent}, while the set of MLUs, wireless channels, and QoS dynamics form the \emph{environment}~\cite{10591707}. At each time slot, the agent observes the current system state, selects an action, and receives a scalar reward. In this MDP, the state summarizes real-time network and device conditions, the action specifies offloading and power allocation decisions, and the reward reflects latency performance and user-perceived quality under compact LLM deployment.

\subsubsection{Action Space}

At each time slot $t$, the agent jointly determines the offloading ratio and transmit power of each MLU. The action space is defined as
\begin{equation}
    \mathcal{A}
    = \big\{\{\alpha_k(t)\}_{k\in\mathcal{K}},\,\{p_k(t)\}_{k\in\mathcal{K}}\big\},
\end{equation}
where $\alpha_k(t)\in[0,1]$ denotes the offloading ratio of MLU $k$, and $p_k(t)\in[0,P_k^{\max}]$ denotes its uplink transmit power.

\subsubsection{State Space}

The state $s_t$ collects information that are relevant for offloading and power control. It includes recent actions, QoS indicators, and latency performance. The state space is defined as
\begin{equation}
\begin{aligned}
    \mathcal{S}
    = \big\{&
      \{\alpha_k(t)\}_{k\in\mathcal{K}},\,\{p_k(t)\}_{k\in\mathcal{K}},\,
      \{\tilde{\mathcal{A}}_k(t)\}_{k\in\mathcal{K}},\\
      &\{\tilde{\mathcal{H}}_k(t)\}_{k\in\mathcal{K}},\,
      \{L_k(t)\}_{k\in\mathcal{K}}
      \big\},
\end{aligned}
\end{equation}
where $\tilde{\mathcal{A}}_k(t)$ and $\tilde{\mathcal{H}}_k(t)$ denote the effective accuracy and hallucination rate of MLU $k$ at time $t$, and $L_k(t)$ is the overall latency defined in~\eqref{time_latency}. In implementation, $s_t$ is formed by concatenating these features across all MLUs, together with exogenous information such as current task sizes and channel gains.

\subsubsection{Reward Function}

To promote low latency while respecting QoS and energy limits, we design a reward that decreases when the aggregate latency or constraint violations increase. The instantaneous reward at time slot $t$ is defined as
\begin{equation}
    r_t = \frac{1}{\sum_{k=1}^{K} L_k(t) + \omega\,\Omega(t)},
\end{equation}
where $\omega>0$ is a penalty coefficient and $\Omega(t)$ captures violations of accuracy, hallucination, and energy constraints, i.e.,
\begin{equation}
\begin{aligned}
\Omega(t) ={}&
\max\!\Big( K \mathcal{A}^{\min} - \sum_{k=1}^{K} \tilde{\mathcal{A}}_k(t),\, 0\Big) \\[3pt]
&+ \max\!\Big(\sum_{k=1}^{K} \tilde{\mathcal{H}}_k(t) - K \mathcal{H}^{\max},\, 0\Big) \\[3pt]
&+ \max\!\Big(\sum_{k=1}^{K} \big(E_k^{\text{local}}(t)+E_k^{\text{off}}(t)\big)
              - \sum_{k=1}^{K} E_k^{\max},\, 0\Big).
\end{aligned}
\end{equation}
Here, $\mathcal{A}^{\min}$ and $\mathcal{H}^{\max}$ are the accuracy and hallucination thresholds in~\eqref{cons_acc}–\eqref{cons_hallu}, respectively. The terms $E_k^{\text{local}}(t)$ and $E_k^{\text{off}}(t)$ denote the local computation and offloading energy of MLU $k$, respectively. $E_k^{\max}$ is its per-slot energy budget. When all QoS and energy constraints are satisfied, $\Omega(t)=0$ and the reward reduces to the inverse of the total latency. When any constraint is violated, $\Omega(t)>0$ reduces the reward, which guides the agent to balance accuracy, hallucination, and energy limits while minimizing latency.

\subsection{Computational Complexity Analysis}
The complete implementation of the proposed world model-PPO algorithm is provided in Algorithm.~\ref{alg:wm_ppo}. {The computational complexity of the proposed world model-PPO algorithm mainly arises from the training processes of the actor-critic neural network and the world model. For the actor-critic network, each forward and backward pass has a complexity of $\mathcal{O}(\sum\nolimits_{l=1}^L n_{l-1} \cdot n_l)$, where $L$ is the number of hidden layers, and $n_l$ is the number of units in the $l$-th layer~\cite{zhang2023energy}. Additionally, the size of the input layer (i.e., $n_0$) corresponds to the state dimension, while the output layer (i.e., $n_l$) size matches the action dimension. The world model is based on the RSSM, which consists of an encoder, prior and posterior networks, a recurrent gated recurrent unit (GRU) cell, and decoders for predicting the next state, reward, and termination. The computational complexity for each RSSM step is $\mathcal{O}(n_h^2 + n_z^2 + n_h \times n_z + n_h \times n_a)$, where $n_z$ is the latent state dimensionality.}

\begin{figure}[t]
\centering{\includegraphics[width=0.43\textwidth]{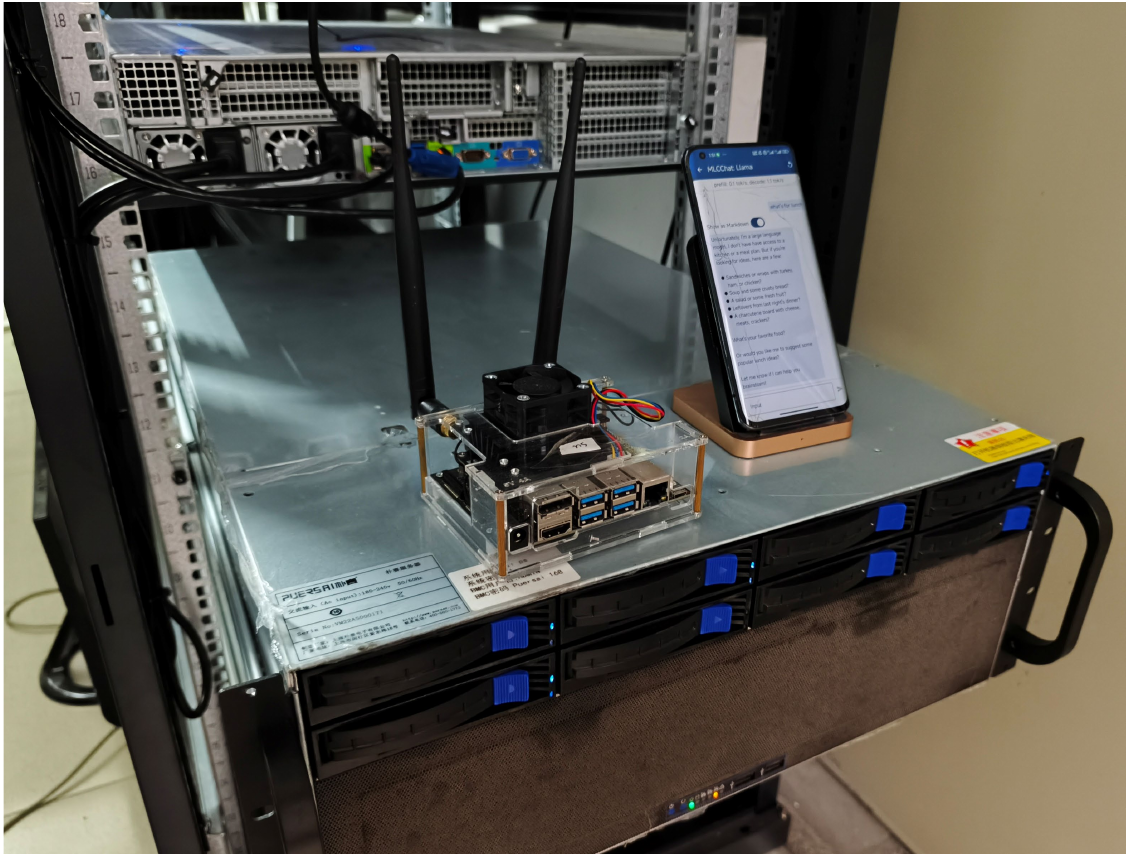}}
\caption{Hardware setup of the compact LLM offloading testbed. The MEC server is equipped with an Intel Xeon Platinum 8380 CPU, an NVIDIA H200 GPU with 128 GB of RAM. The user devices include: a) An Nvidia Jetson Nano B01 Developer Kit 4 GB with dual antennas for wireless communications and a 128-core Maxwell GPU. b) A Xiaomi 10 Ultra smartphone with 12GB of RAM and a Qualcomm Snapdragon 865 processor.}\label{device_setup}
\end{figure}


\section{Experimental Setup and Results}

\subsection{Experimental Setup}

\subsubsection{System Parameters}
We consider an MEC system with $K=2$ MLUs randomly located in a circular area of radius $20$\,m, while the MEC server is placed at the center of the cell. {The reference channel gain $g_0$ is set to $-30$ dB~\cite{lv2024ris}.} Unless otherwise stated, the maximum uplink transmit power of each MLU is set to $P_k^{\max}=33$\,\text{dBm}, and the receiver noise power is set to $\sigma^2=-104$\,\text{dBm}. The wireless channel follows the Rician fading model in Section~III-B with Rician factor $\kappa=8$, and all transmissions share an uplink bandwidth of $B=10$\,\text{MHz}~\cite{10771695}. For computation, each MLU is equipped with a local CPU of clock frequency
$f_k = 2\times 10^{9}$\,\text{cycles/s} (i.e., $2$\,GHz), while the MEC server has a total computational capability of
$F = 10\times 10^{9}$\,\text{cycles/s} (i.e., $10$\,GHz). The LLM inference workload is characterized by
$\phi = 900$ CPU cycles per processed bit, which is used consistently for both local and edge-side computation. The per-slot energy budget of each MLU is set to $E_k^{\max}=2$\,\text{J}, which limits the combined energy consumption of local computation and uplink transmission. The computation energy coefficient in the local energy model is set to $\kappa_{\text{comp}} = 10^{-28}$ {J/cycle}.

\subsubsection{ECLD Experimental Setup, Datasets, and Metrics}

We evaluate the proposed ECLD framework from both a system and a model perspective. All training and compression procedures are implemented in PyTorch and executed on a workstation equipped with an NVIDIA H200 GPU and an Intel Xeon Platinum 8380 CPU, which is also used to train the compact LLMs and the world model-PPO algorithm, as illustrated in Fig.~\ref{device_setup}. For on-device deployment, we consider two representative edge platforms. The first is an NVIDIA Jetson Nano B01 Developer Kit with 4\,GB RAM, dual antennas for wireless communication, and a 128-core Maxwell GPU. The second is a Xiaomi 10 Ultra smartphone with 12\,GB RAM and a Qualcomm Snapdragon 865 processor. These devices represent typical low-power edge hardware and high-end mobile terminals, and allow us to assess the practicality of compact LLM deployment under different resource profiles. We benchmark our approach using several mainstream foundation models, including Qwen3-8B, Llama-3.1-8B, and Mistral-12B. For each base model, we construct multiple compact variants using low-bit quantization, pruning, knowledge distillation, and the full ECLD pipeline. These variants are compared in terms of both efficiency and generation quality, as detailed below.

\textbf{Accessibility:}
To evaluate deployability on resource-constrained devices, we measure \emph{Accessibility} as the on-disk size of the model parameters. This metric reflects how effectively each compression strategy reduces storage overhead and improves suitability for on-device or distributed deployment.

\textbf{Energy Consumption:}
To quantify operational efficiency during inference, we measure \emph{Energy Consumption} as the average energy required to generate a complete response to a single query. This provides a hardware-agnostic indicator of the inference-time energy footprint and shows how model compactness translates into sustainable usage in practice.

\textbf{Accuracy and Hallucination Rate:}
We evaluate the impact of compression and ECLD on factual accuracy using the answer matching criterion defined in Equation~\eqref{eq:acc} on the WebQuestionsSP dataset~\cite{yih2016value}. Additionally, we assess the effect of compression and ECLD on hallucination rates using selfcheckgpt~\cite{manakul2023selfcheckgpt} with the sentence-level scoring criterion specified in Equation~\eqref{eq:hallu}.

\subsubsection{World model-PPO Experimental Setup}
{Both the actor and critic networks in PPO are implemented as two-layer multilayer perceptrons (MLPs), each consisting of $256$ hidden units per layer. The learning rate for both networks is set to $10^{-5}$. The PPO clipping parameter $\epsilon$ is set to $0.1$, the discount factor $\gamma$ is set to $0.99$, and the policy is updated for $10$ epochs per iteration. As for the world-model, the MLP components of the RSSM also adopt $256$, the learning rate for updating the RSSM is $10^{-3}$, the number of imagined steps $H$ is set to $3$, the contribution coefficient of imaginary TD $\lambda_{\rm{wm}}$ is set to $0.5$, the entropy coefficient $\beta_{\rm{ent}}$ is set to $0.001$, and the auxiliary imagination loss contribution coefficient $\eta$ is set to $0.3$. In addition, both the reward scaling factor and the KL-divergence regularization coefficient are set to $1$. The minimum accuracy threshold $\mathcal{A}^{\min}$ is set to $0.6$, and the maximum hallucination threshold $\mathcal{H}^{\max}$ is set to $0.78$~\cite{wang2025world,zhao2025world}.}

\begin{table*}[t]
\centering
\caption{Performance comparison across three LLMs under different compact methods}
\begin{tabular}{cccccc}
\toprule
\textbf{Model} & \textbf{Method} & \textbf{Hallucination (↓)} & \textbf{Accuracy (\%, ↑)} & \textbf{Accessibility (MB, ↓)} & \textbf{Energy Consumption (W·h, ↓)} \\
\midrule

\multirow{5}{*}{\raisebox{-6pt}{\textbf{Llama3.1-8B}}} 
& Original & 0.77 & 70.30 & 15316.53 & 0.24 \\ 
& Quantization~\cite{xiao2023smoothquant} & 0.82 & 24.99 & 4308.13 & 0.13 \\ 
& Pruning~\cite{ma2023llm} & 0.70 & 30.76 & 13236.46 & 0.27 \\ 
& Pruning + Distillation~\cite{muralidharan2024compact} & 0.85 & 62.11 & 13236.46 & 0.20 \\ 
& Ours & \textbf{0.65} & 59.05 & \textbf{3,336.18} & \textbf{0.12} \\
\midrule

\multirow{5}{*}{\raisebox{-6pt}{\textbf{Qwen3-8B}}} 
& Original & 0.77 & 68.21 & 15623.45 & 0.17 \\ 
& Quantization~\cite{xiao2023smoothquant} & 0.78 & 67.51 & 5792.20 & 0.13 \\ 
& Pruning~\cite{ma2023llm} & 0.66 & 19.32 & 11206.87 & 0.11 \\ 
& Pruning + Distillation~\cite{muralidharan2024compact} & 0.95 & 37.01 & 11206.87 & 0.11 \\ 
& Ours & 0.97 & 36.35 & \textbf{4,652.80} & \textbf{0.09} \\
\midrule

\multirow{5}{*}{\raisebox{-6pt}{\textbf{Mistral-12B}}} 
& Original & 0.52 & 65.95 & 23360.83 & 0.35 \\ 
& Quantization~\cite{xiao2023smoothquant} & 0.94 & 64.94 & 7,926.37 & 0.29 \\ 
& Pruning~\cite{ma2023llm} & 0.80 & 20.71 & 16,048.67 & 0.32 \\ 
& Pruning + Distillation~\cite{muralidharan2024compact} & 0.84 & 75.79 & 16,048.67 & 0.26 \\ 
& Ours & 0.87 & 71.88 & \textbf{5,660.25} & \textbf{0.24} \\
\bottomrule

\multicolumn{6}{p{0.98\textwidth}}{\footnotesize\textbf{(a) Hallucination rate (↓)}, where a lower value indicates better factual consistency and reliability;}\\
\multicolumn{6}{p{0.98\textwidth}}{\footnotesize\textbf{(b) Generalization accuracy (↑)}, where a higher value reflects stronger task adaptability and reasoning capability;}\\
\multicolumn{6}{p{0.98\textwidth}}{\footnotesize\textbf{(c) Accessibility (↓)} represents the model storage footprint, where a lower value denotes smaller on-device storage requirements and improved deployability;}\\
\multicolumn{6}{p{0.98\textwidth}}{\footnotesize\textbf{(d) Energy consumption (↓)}, where a lower value indicates better energy efficiency and runtime sustainability.}

\end{tabular}
\label{tab:llm_compression_comparison}
\end{table*}


\subsection{Experimental Results}


\subsubsection{ECLD Model Efficiency Analysis}

To evaluate the practical deployability of different compact–model construction strategies, we compare ECLD against pure quantization, pruning, and pruning+distillation baselines across three representative backbones (i.e., Llama-3.1-8B, Qwen-3-8B, and Mistral-12B). The results in Table~\ref{tab:llm_compression_comparison} show the hallucination rate, accuracy, model accessibility, and energy consumption for each method. In terms of \emph{accessibility} (on-disk model size), the proposed ECLD pipeline consistently yields the most compact models. For example, the Llama-3.1-8B variant produced by ECLD is reduced from about $15.3$\,GB to $3.3$\,GB, and the Qwen-3-8B and Mistral-12B models are compressed from roughly $15.6$\,GB and $23.4$\,GB to $4.7$\,GB and $5.7$\,GB, respectively. This corresponds to a storage reduction of about $70$–$80\%$, which is substantially larger than that achieved by pure quantization or pruning. The proposed pipeline therefore enables deployment on devices that cannot host even an 8-bit quantized baseline. ECLD also provides favorable energy–quality tradeoffs. For Llama-3.1-8B, ECLD halves the energy consumption (from $0.24$\,Wh to $0.12$\,Wh) while keeping accuracy within about $10$ percentage points of the full model and reducing the hallucination rate from $0.77$ to $0.65$. For Qwen-3-8B and Mistral-12B, the ECLD variants cut energy usage by roughly $30$–$50\%$ compared with the original models, and maintain higher accuracy than simple pruning while using much smaller storage than pruning+distillation. Although aggressive compression can slightly increase hallucination in some cases (e.g., Qwen-3-8B), the overall balance across hallucination, accuracy, accessibility, and energy consumption is consistently better than that of the baselines. These results show that the proposed ECLD framework is not only able to compress large foundation models into edge-deployable forms, but also preserves a desirable level of reasoning quality and energy efficiency. The ECLD variants therefore serve as strong compact LLM candidates for the subsequent MEC offloading experiments.

\subsubsection{Performance Analysis of ECLD Reasoning Results}

\begin{figure*}[t]
\centering{\includegraphics[width=\textwidth]{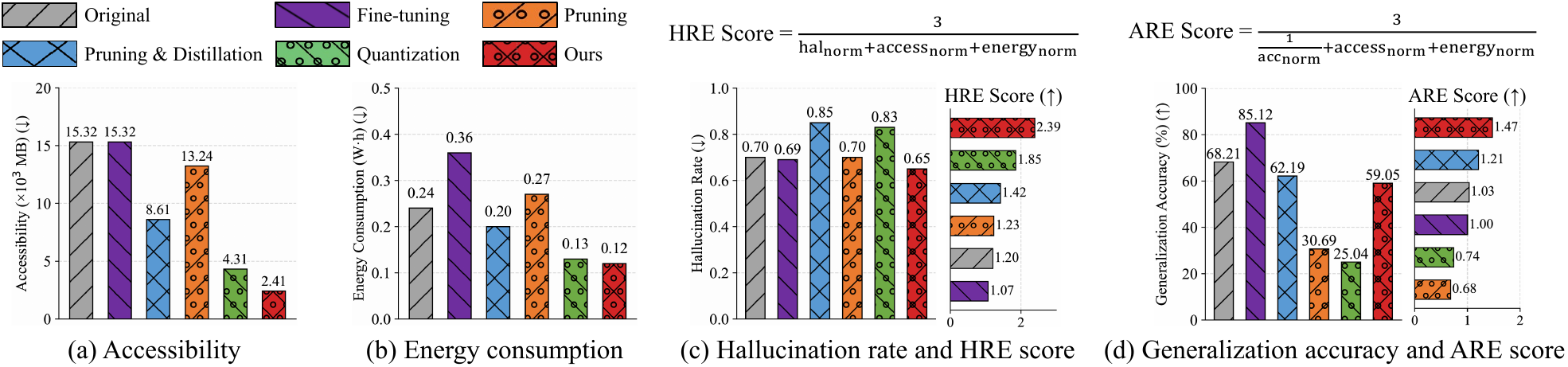}}
\caption{Performance comparison of four compact LLM strategies and the proposed approach across four key evaluation metrics tested on Llama3.1-8B model. 
The hallucination-aware resource efficiency (HRE) Score and accuracy-aware resource efficiency (ARE) Score are defined as the harmonic mean of the max-normalized accessibility, energy consumption, and either hallucination rate (for HRE) or generalization accuracy (for ARE).} \label{fig_strategy}
\end{figure*}

To assess the reasoning capability and factual reliability of the compact models, we conduct accuracy–oriented and hallucination–oriented evaluations on the WebQuestionsSP and WikiBio GPT-3 Hallucination datasets. As shown in Table~\ref{tab:llm_compression_comparison}, the models obtained by ECLD consistently achieve higher accuracy than those compressed only by quantization or pruning, while using much smaller model sizes. For the Llama-3.1-8B backbone, for example, ECLD reduces the model size from about $15$\,GB to $3.3$\,GB and still maintains an accuracy of $59.05\%$, which is close to the original model (i.e., $68.21\%$) and clearly higher than the quantized and pruned baselines. Figure~\ref{fig_strategy} further illustrates the quality–efficiency tradeoff. As shown in Fig.~\ref{fig_strategy}(c), the ECLD variant attains the lowest hallucination rate among all compact methods, and this directly leads to the highest Hallucination-aware Resource Efficiency (HRE) score once accessibility and energy consumption are normalized and aggregated. Similarly, Fig.~\ref{fig_strategy}(d) shows that ECLD preserves high generalization accuracy while operating with low storage and energy cost, which results in the largest accuracy-aware resource efficiency (ARE) score. 

\begin{figure*}[h]
\centering{\includegraphics[width=\textwidth]{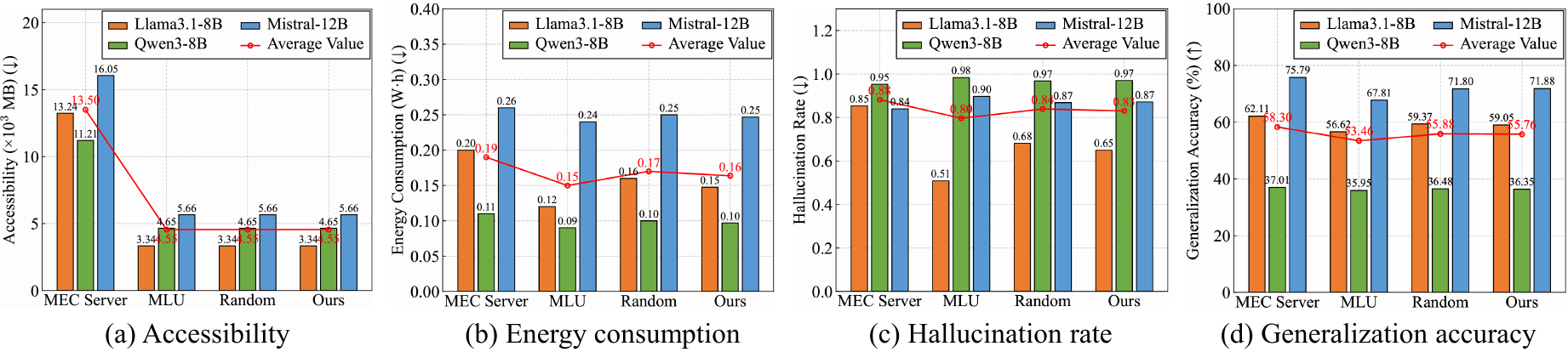}}
\caption{Comparison of four offloading strategies across three LLMs (i.e., Llama3.1-8B, Qwen3-8B, and Mistral-12B). The red solid line depicts the averaged performance of each offloading strategy under the three models.} \label{fig_offloading}
\end{figure*}

\subsubsection{ECLD Energy Consumption Analysis}
To evaluate the operational efficiency of different compact–model construction strategies, we compare their average inference–time energy consumption. As shown in Table~\ref{tab:llm_compression_comparison} and Fig.~\ref{fig_strategy}(b), the models produced by ECLD require the least energy on average among all compact baselines. Across the three backbones (Llama-3.1-8B, Qwen3-8B, and Mistral-12B), ECLD reduces the per-query energy by roughly half compared with the original full models, and also improves over pure pruning and pruning–plus–distillation schemes. Compared with isolated low-bit quantization, ECLD achieves similar or lower energy consumption while avoiding the sharp accuracy and hallucination degradation observed for heavily quantized models. These gains stem from the joint design of pruning, distillation, and quantization in ECLD. The resulting compact LLMs therefore execute each forward pass with both fewer operations and better calibrated internal representations, leading to consistently lower energy consumption during generation. The reductions in energy consumption at the model level lower the cost of long-term deployment and ultimately enlarge the set of devices on which LLMs can run.

\subsubsection{Performance of Inference Offloading Analysis} Fig.~\ref{fig_offloading} demonstrates that the proposed world-model-assisted inference offloading strategy consistently achieves the best overall quality--efficiency tradeoff across all three compact LLMs. Although compact-model deployment inherently improves accessibility and reduces energy consumption compared with MEC-server execution, different offloading policies lead to substantially different inference outcomes. Compared with the MLU and random baselines, the proposed scheme attains the lowest average hallucination rate while maintaining the highest generalization accuracy, without increasing resource consumption. This improvement stems from the predictive capability of the world model, which estimates the long-term impact of offloading decisions on both resource utilization and inference quality. Rather than making myopic decisions based solely on current system conditions, the proposed strategy proactively selects execution locations that better preserve the reasoning capability of compact LLMs while maintaining efficient resource usage. These results suggest that intelligent offloading can effectively compensate for the performance loss introduced by model compression, thereby enabling high-quality LLM inference in resource-constrained edge environments.

\subsubsection{Convergence Behaviour of World model-PPO}
\begin{figure}[t]
    \centering
    \includegraphics[width=0.43\textwidth]{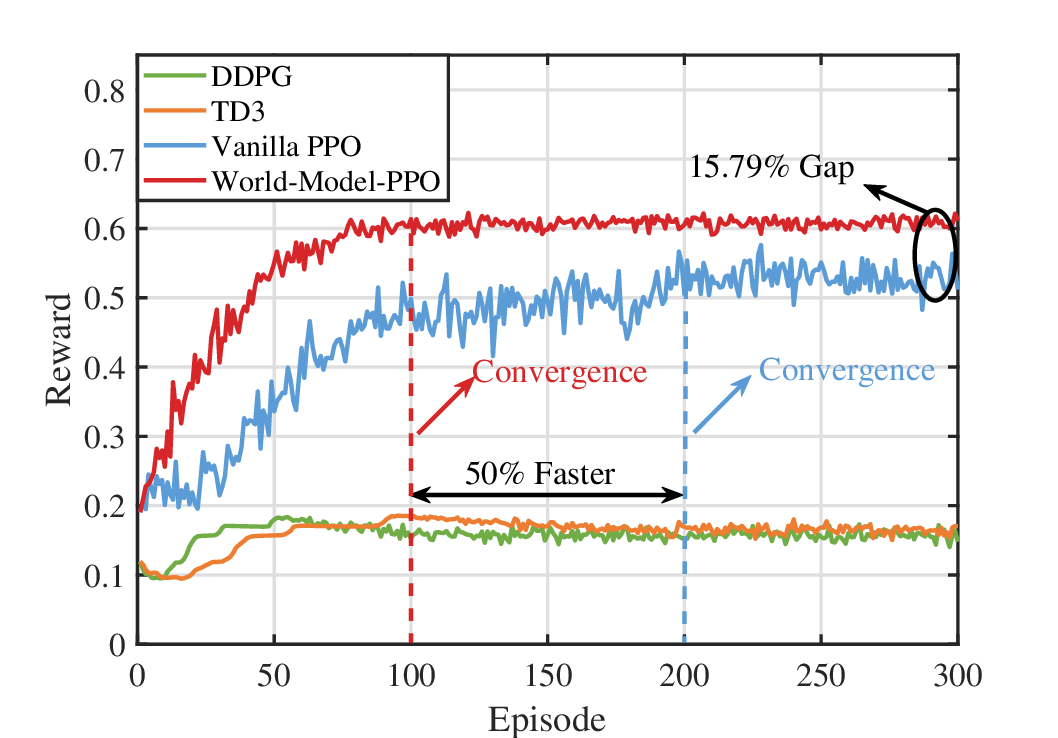}
    \caption{Convergence behaviour of the proposed world model-enhanced PPO and baselines.}
    \label{fig:rl_convergence}
\end{figure}

After deploying compact LLMs with ECLD, we turn to dynamic offloading and power control using the proposed world model-PPO algorithm. Fig.~\ref{fig:rl_convergence} compares the convergence behaviours of four baselines, namely TD3, DDPG, vanilla PPO, and world model-PPO, under the same system settings and reward definition. As training progresses, world model-PPO achieves a much steeper reward increase in the early stage and reaches a stable plateau at around $100$ episodes, whereas vanilla PPO requires roughly $200$ episodes to converge. This corresponds to about $50\%$ faster convergence for world model-PPO. In addition, the final converged reward of world model-PPO is about $15.8\%$ higher than that of vanilla PPO, while TD3 and DDPG saturate at significantly lower reward levels with larger fluctuations. These results indicate that world model-PPO not only learns a better offloading and power-control policy, but also does so with improved sample efficiency. The performance gain comes from the integration of the world model into PPO. The one-step world model-boosted value targets reduce variance in critic learning, and the short imagination rollouts provide an additional, long-horizon learning signal for the actor. As a result, world model-PPO can anticipate the impact of current actions on future latency, energy, and QoS, instead of reacting only to immediate feedback. 

\subsubsection{Scalability with Respect to the Number of MLUs and Task Size}

\begin{figure}[t]
    \centering
    \includegraphics[width=0.44\textwidth]{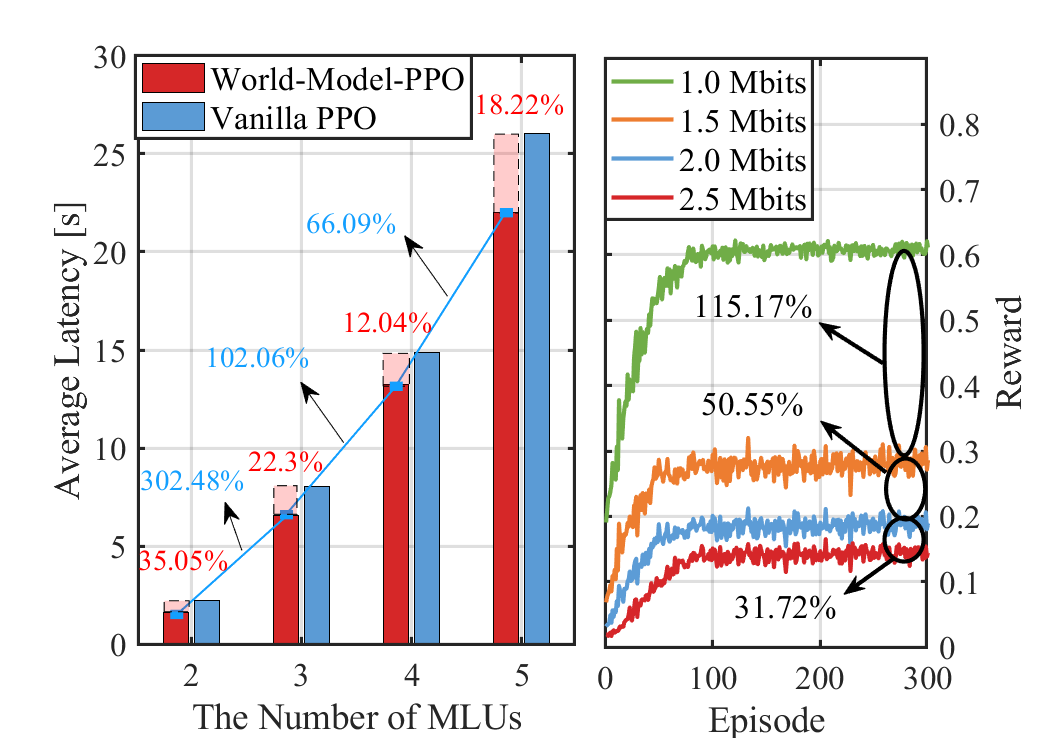}
    \caption{(a) Average inference latency versus the number of MLUs. (b) Reward convergence of world model-PPO under different task sizes.}
    \label{fig:latency_mlu}
\end{figure}

After validating the convergence behavior of world model-PPO, we further examine how its performance scales with the number of MLUs and with different task sizes, as illustrated in Fig.~\ref{fig:latency_mlu}. In Fig.~\ref{fig:latency_mlu}(a), the average latency increases for both methods as the number of MLUs grows from $2$ to $5$, which reflects the stronger uplink interference and the heavier load on the MEC server. For every operating point, however, world model-PPO achieves a lower latency than vanilla PPO. The latency reduction ranges from about $12\%$ to more than $30\%$, showing that the world model helps the algorithm coordinate offloading ratios and transmit powers across users more effectively when their decisions are strongly coupled through shared spectrum and edge computation resources. Fig.~\ref{fig:latency_mlu}(b) shows the reward evolution of world model-PPO under different packet sizes, from $1.0$~Mbits to $2.5$~Mbits. Smaller packets yield higher steady-state rewards and faster convergence, since they require fewer CPU cycles and less uplink transmission time for each LLM query, which leads to lower latency and energy consumption. As the packet size increases, the system becomes more heavily loaded, the achievable reward decreases, and the learning process slows down. Nevertheless, world model-PPO still converges to a stable policy in all cases, indicating that the world model-PPO can adapt its offloading and power decisions to a wide range of input sizes. This behavior is consistent with the characteristics of LLM inference, where longer prompts and responses translate into larger data volumes and higher computation costs.

\begin{figure}[t]
    \centering
    \includegraphics[width=0.43\textwidth]{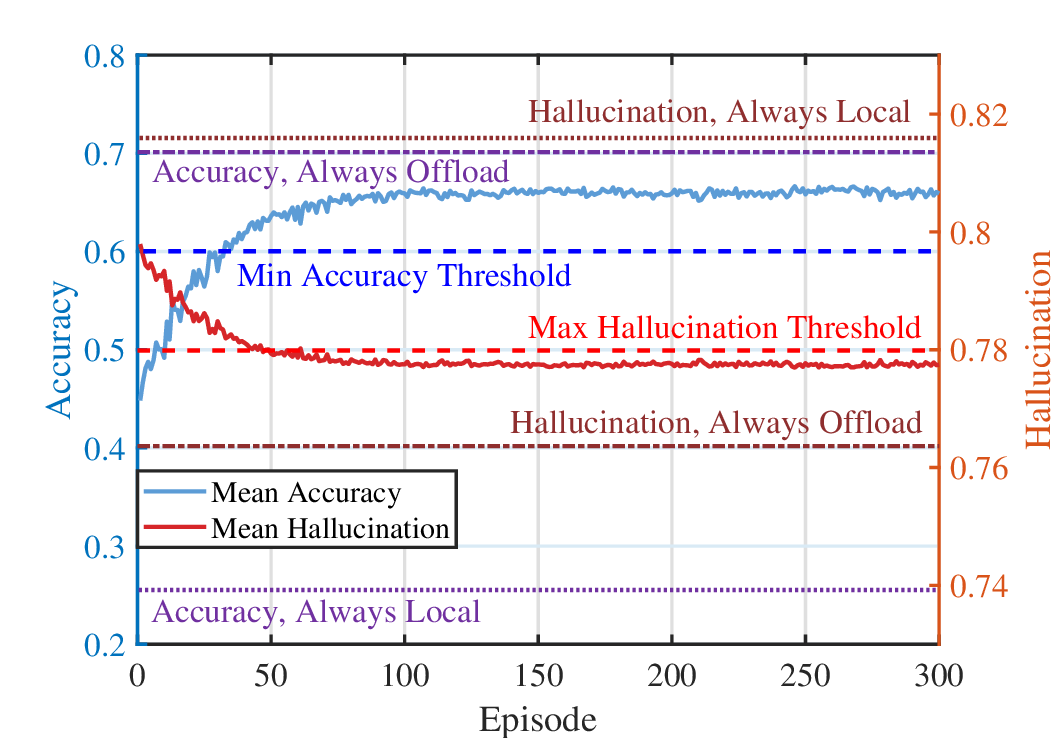}
    \caption{Accuracy and hallucination rates under dynamic offloading, always-local, and always-offload.}
    \label{fig:accuracy_hallu}
\end{figure}

\subsubsection{Accuracy and Hallucination under Dynamic Offloading, Always-Local, and Always-Offload}

Fig.~\ref{fig:accuracy_hallu} shows the learning curves of the dynamic offloading policy learned by world model-PPO in terms of mean accuracy and mean hallucination. For comparison, we also show two static baselines, namely the always-local and always-offload policies, together with the target thresholds $\mathcal{A}^{\min}$ and $\mathcal{H}^{\max}$. As training proceeds, the accuracy of the dynamic policy quickly rises above the always-local baseline and approaches the always-offload upper bound, while consistently remaining higher than $\mathcal{A}^{\min}$. In contrast, its hallucination rate drops below the always-local level and stays only slightly above the always-offload lower bound, yet remains safely under $\mathcal{H}^{\max}$. These trends indicate that world model-PPO learns to offload primarily when the local compact LLM is likely to violate the QoS targets, and otherwise favors local execution to reduce latency and energy consumption. Consequently, the dynamic offloading strategy achieves a balanced operating regime between the two static extremes, delivering near–cloud-level generation quality while preserving the efficiency advantages of on-device inference.

\section{Conclusion}
This paper has investigated compact LLM deployment and world-model-assisted inference offloading for MEC. We have proposed the ECLD framework that jointly performs structured pruning, knowledge distillation, and low-bit quantization to obtain edge-deployable LLM variants under diverse device capabilities. We have further formulated a long-term latency minimization problem with energy, accuracy, and hallucination constraints, and developed a world model-PPO algorithm that leverages a recurrent world model for value boosting and imagination-based policy improvement. Simulation results have demonstrated that ECLD can substantially reduce model size and energy consumption while largely preserving or even improving hallucination behavior, and that world model-PPO can achieve lower latency and faster convergence than baseline offloading schemes, while satisfying LLM-specific QoS requirements.

\bibliographystyle{ieeetr}
\bibliography{ref}

\end{CJK}
\end{document}